# Supernovae Light Curves: An Argument for a New Distance Modulus

Jerry W. Jensen                    April 6, 2004

## Abstract

Supernovae Ia (SNe Ia) light curves have been used to prove the universe is expanding. As standard candles, SNe Ia appear to indicate the rate of expansion has increased in the past and is now decreasing. This independent evaluation of SNe Ia light curves demonstrates a Malmquist Type II bias exists in the body of supernova data. If this bias is properly addressed, there is very little budget for time dilation in the light curves of supernova. A non-relativistic distance modulus is proposed, which is based on the predictable attenuation of light by an intergalactic CREIL (Coherent Raman Effects on Incoherent Light) radiation transfer functions.

CREIL intrinsically displace the red shift of quasars and other blue objects relative to the rest of the universe. Evidence of these radiation transfer functions is found in the spectral distribution of quasars, and in the continuum emission and total luminosity functions of the universe in the IR and NIR bandwidths. These radiation transfer functions may also contribute to the Tully-Fisher relationship, and resolve the degeneracy in the microwave power function. The anomalous accelerations of the Pioneer 10 and 11 space probes towards the sun may be explained by these mechanisms.

Without relativistic expansion, new explanations are required to explain background radiation and light element synthesis. An electro-gravimetric energy transfer function activated at extremely low temperatures in molecular hydrogen by a Bose-Einstein condensation is proposed as a candidate mechanism. This proposed solution to Obler's paradox hints of a tantalizing link between the large numbers theories of Mach and Dirac and a non-Friedman field limited gravimetric tensor. This is a highly speculative reintroduction of Hoyle's steady-state cosmology.

## Introduction

For most of the 20[th] century astronomical observations such as galactic evolution, heavy metal abundance, supernovae light curves and the cosmic microwave background have fallen within the constraints of the Einstein - de Sitter Big Bang model. In the last two decades these relationships have become severely strained. The universe is too big and too old; the magnitudes of supernovae are dimming too fast. There are too many radio point sources. The far infrared continuum emissions imply a dusty past that is completely at odds with multi-colored supernovae and quasar spectrums. There is too little anisotropy in the cosmic microwave background to support the observed galaxy super-cluster structure. Heavy metal ratios equal to solar concentrations have been quantified in the most redshifted objects we observe. Something is wrong with this mature theory: It has failed in too many predictions.

To address these shortcomings, the 2003 data release of the Wilkinson Microwave Anisotropy Probe is packaged with more than a dozen papers detailing methodology and cosmological issues. These interpretative dialogues rely on the microwave background and supernovae Ia data to constrain the universe into an extremely tight model. To fit current observations into a modified Big Bang universe, unphysical *ad hoc* parameters are used, inserting copious quantities of 'dark matter' and 'dark energy'. This is not good science.

Critical to the 'WMAP Einstein deSitter' cosmology are the supernova interpretive studies, which predict the rate of expansion the universe is increasing. This paper contends the multicolored light-curves of supernovae, and other statistical evidences indicate a magnitude bias is responsible for most light curve variance, rather than time dilation. A reassessment of fundamental cosmological parameters is in order. This paper touches on a very broad range of astrophysical topics. This is necessary because the complexities of the current model are deeply woven into the fabric of the universe and must be carefully picked apart, string-by-string.

**Part I:** Argues all observed supernova Ia light curves fall within local constraints for binary SN Ia scenarios, regardless of redshift. Statistical indicators are extracted from multi-color light curves and other supernova Ia studies. These demonstrate that there is a selection effect in the magnitude of SNe Ia observed at increasing redshift distances. This bias leaves little residual light curve budget for time dilation, a conclusion supported in the power distribution functions of gamma rays and quasars.

**Part II:** Examines the implication of removing time dilation from the relativistic Hubble magnitude and distance plots, which actually displaces supernovae Ia magnitude functions further from expected values. Here it is demonstrated that a regressive series of natural log functions provides a better fit to the supernovae Ia magnitude plots than strictly relativistic functions.

**Part III:** Shows how coherent radiation transfer functions can be used in models to create the discrete spectral shifts and attenuation factors observed in the universe. Equations that can be derived from radiation transfer functions are proposed which are consistent with the observed attenuations of supernovae at great distances.

**Part IV:** Restates the coherent radiation transfer process described by Moret-Bailly, (CREIL), which is responsible for displacing the redshift of quasars-like objects, *and their host galaxies*, relative to the rest of the universe.

**Part V:** Catalogues a wide variety of cosmic observations that are supportive of a universe with radiation transfer functions in the interstellar medium. The symmetry of the Tully-Fisher relationship, and potential causal connections between quasar spectral properties and very red objects are examined.

**Part VI:** Resolves some of the many problems that arise in the extrapolation of a dynamic-steady-state universe. These include galactic evolution, Periodic quasar epics, Obler's paradox, the Cosmic Microwave Background, and light element synthesis.

**Part VII:** Is a highly speculative attempt to frame a steady state cosmology without invoking a cosmological constant, substituting instead field-limited gravity. A CREIL mechanism is also proposed for the celestial radiation transfer functions necessary in the balance of this paper.

Although this model assumes non-relativistic causality for most redshifting, it does not eschew general relativity, with the sole exception of a Maxwellian coupling of 'gravity' to atoms in Planck increments, thus limiting the field effects of gravity, rather than invoking a cosmological constant. As in the big bang, point violations of the second law of thermal dynamics are required for the universe to exist. This skeletal model assumes a Lorenz invariant universe consistent with general relativity, and that the Cosmological Principle applies.

# Part I: Supernova Ia

Of celestial events, the explosions of supernovae are among the rarest and most spectacular. A supernova Ia (SN Ia) is thought to be the death throws of a white dwarf star depleted of light, fusionable elements. A supernova has accreted enough mass to exceed the Chandrasekhar limit, meaning all baryonic matter is compacted into a critical state. The remaining fusionable matter, consisting primarily of $^{12}$carbon and $^{16}$oxygen, complete a thermonuclear sequence into $^{28}$silica, $^{56}$nickle, $^{56}$cobalt, and finally $^{56}$iron, while the nova explodes in a final, brilliant nuclear spasm. (Arny). The unique spectral fingerprint, and narrow range of magnitude of this brilliant explosion makes SN Ia excellent 'standard candles' for measuring distances.

Characteristic of supernovae Ia are the shapes of the light-curves over many days, both approaching and receding from maximum luminescence (Fig 1). In 1939, Wilson hypothesized that if these supernovae could be identified at high Doppler shifts (redshifts), the time dilation characteristics of an expanding universe could be verified by plotting the light-curves of supernovae that exploded eons ago, and comparing them with 'local' supernovae (Leibundgut).

For example, assume a local supernova Ia (SN Ia) dims by one magnitude from its peak magnitude in fifteen days. Now assume an *exactly identical* supernova is moving away from our galaxy at about half the speed of light (redshift $\cong 0.5$). In an expanding universe, this twin supernova would *appear* to explode more slowly, losing one magnitude in 22.5 days {(1+0.5) x 15 days}.

Current interpretations of supernovae Ia light curves at high redshift distances support this thesis. However, as the database of supernovae light-curves expands, it has become evident that even the most carefully defined SN Ia exhibit light-curves that vary in absolute magnitude (**Figure 1**). Even though the variance in magnitudes is relatively small, the light-curves (in days) vary significantly. Longer light-curves correlate with higher magnitude SNe Ia. (Goldhaber 1996, Hamuy 1996.)

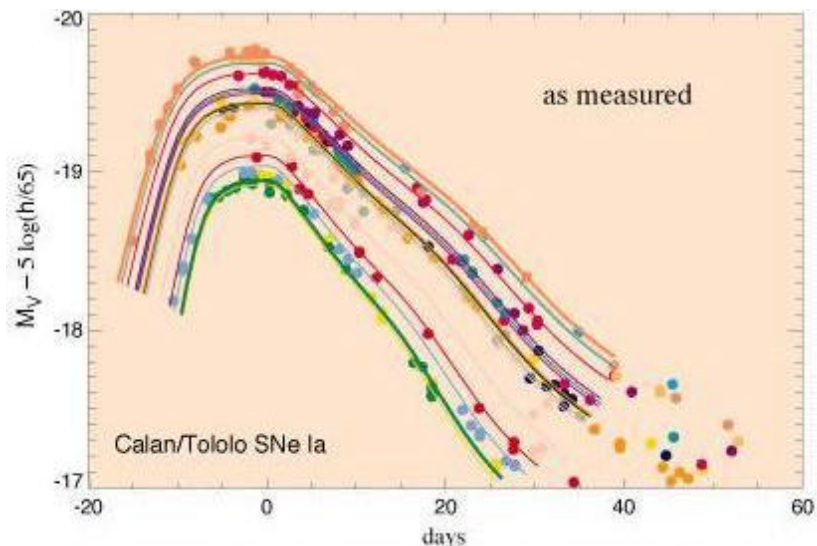

**Figure 1** (SCP). Light curves over days of even the most carefully defined Sne IA vary in both absolute magnitude and width. Higher magnitude supernovae tend to have longer light curves (in days).

To use supernovae as standard candles, a 'stretch factor' method has been developed: First the light-curves are converted to a common epoch using 'cross correlated K factors' to normalize the spectra. Then the light-curve width is adjusted for redshift time dilation, *finally* a free parameter, the stretch factor, is used to scale both the magnitude and the width of the curve to match a template in the common era (Perlmutter). Using this stretch factor method to determined Supernova Ia absolute magnitude, researches have concluded cosmic expansion is occurring with a confidence as high as *twenty-one* Sigma (Nugent). Supernova researches have also used this data to conclude the rate of expansion of the universe was once increasing (Riess, **Note 1**).

### The Stretch Factor Method: No Budget for Malmquist effects

The 'Stretch factor method relies upon a free parameter to associate the magnitude of the light curve with the light curve width. There are at least four potential flaws in these analyses: 1) *A correction is made for redshift before other potentially distance biasing effects are addressed. 2) The sample is normalize at a reference value midway between local and distant supernova. Any linear distance dependencies are buried by this technique. 3) The 'free parameter' is unique to each supernova and assumes no lateral variance in origin. 4) No allowance is made for a Malmquist* type II *bias* (Teerikorpi).

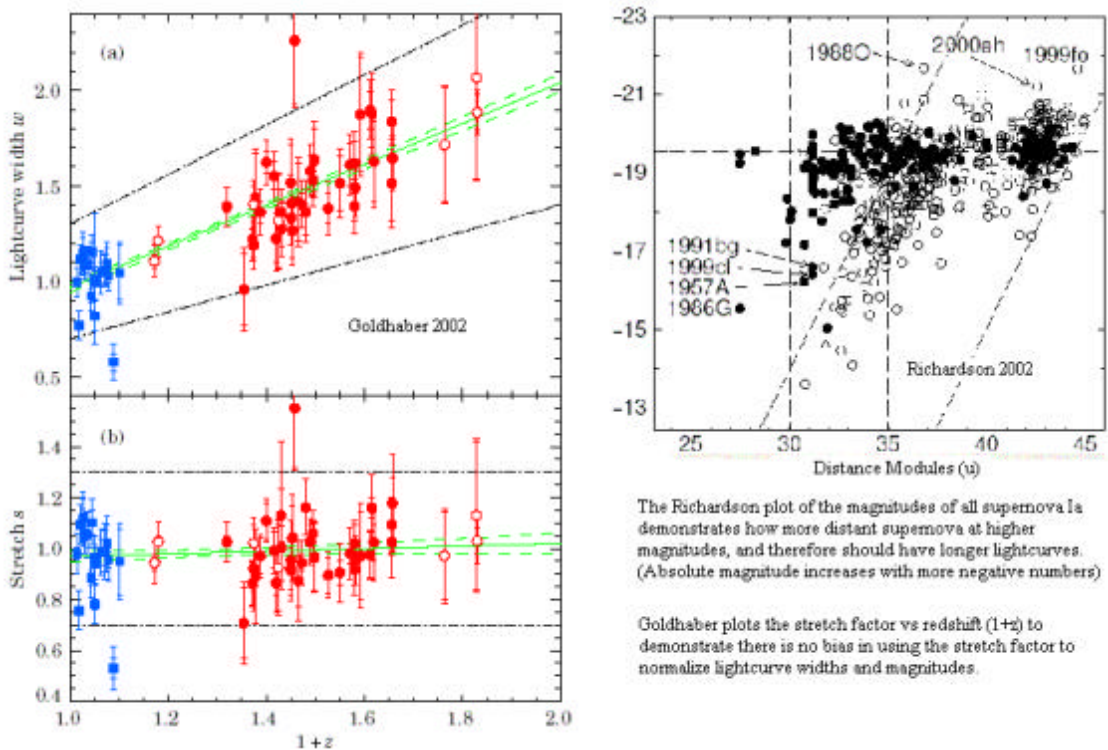

**Figure 2** (Goldhaber). *Goldhaber* makes a case very good for time dilation using the stretch factor developed in *Perlmutter 1998*, demonstrating no residues, but it does not take into account the potential magnitude bias in more distant Supernovae Ia as charted in *Richardson.*.

**Figure 3** (Richardson). Notice how the highest magnitude SNe Ia are located at greater distances than the weakest magnitude SNe Ia. The fact that *Goldhaber's* test for Malmquist bias does not detect this systemic trend demonstrates an error may exist in the treatment of light-curves by *Goldhaber*.

By correcting for time dilation *before* selecting a 'template' for curve matching, *Goldhaber* may underestimate the true magnitude. If the Malmquist bias is proportional to the redshift, and it should be, this method of analysis rolls this magnitude bias right into the time dilation factor.

**Color Magnitude Intercept Calibration: A Time-independent Supernova Size Indicator**
In a paper introducing the Color Magnitude Intercept Calibration (CMAGIC) methodology,
Wang, Goldhaber, Aldering and Perlmutter observe:

The K-corrected fits of the *B* verses *B-V* Color curves show slopes that are consistently smaller than those
without K corrections…The data suggest strongly that the assumption of a single parameter family is not
appropriate in describing the variations of the BV values.

While this possible bimodality in the light-curves of SN Ia located at high redshift may indicate a
puzzling evolutionary branching, it could also be an indication time dilation corrections skew the
data, obscuring a Malmquist type II bias. It could also mean the most distant supernovae Ia
included in the study are like local *hypernovae*.

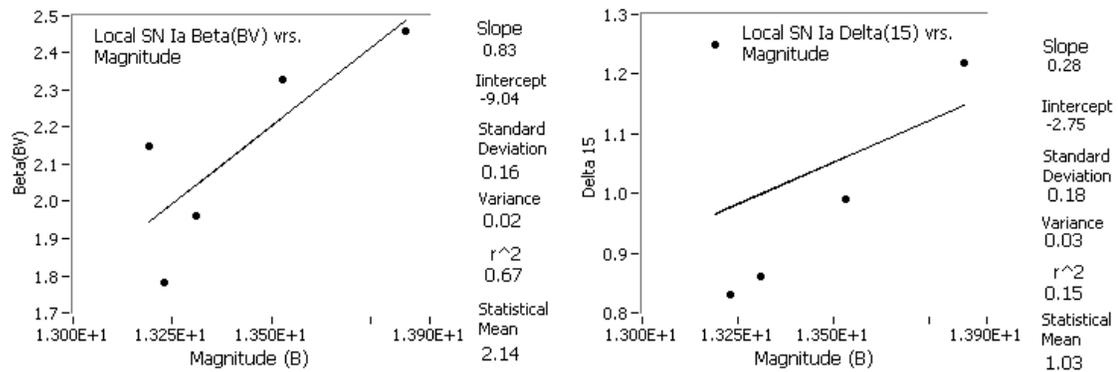

**Figure 4** (left). Local supernova Blue Magnitude vs. Beta(BV). **Figure 5** (right). The local Delta(15)
vs. Blue Magnitude function. (Figures 4-9 use data tabulated in Wang Multi-Color, 2003).

Figures 4 - 9 demonstrate how the multi-colored light curve methodology suggests a time dilation
biasing error in the data reduction of supernova light curves. Beta*(BV)* is defined in *Wang* as an
intrinsic indicator of magnitude, based on the difference between blue and visible magnitudes in a
linear section of the supernovae Ia light curves. Since there is no derivative with respect to time,
the only time dilation factors are minimal band adjustments in the k corrections. In the small local
sample (figure 4), the trend line indicates weaker supernovae have greater Beta(*BV*) values. This
trend is also apparent in local Delta(15) values (figure 5), which are distilled from light curve
widths. This is consistent with the observations of Hamuy and others that intrinsically brighter
SN Ia have longer light curves and correspondingly smaller Delta(15) values. Larger Delta(15)
and larger Beta(*BV*) values both indicate smaller light curves.

If time dilation is a property of the universe, light curve widths must be corrected for time dilation
before comparing delta(15) values. However, if there are non-Doppler causes of redshifting, the
light curve corrections will cause distant supernova to *appear* to have smaller light curves.

Figure 6 and Figure 7 demonstrate an *anti-correlation* between Beta(*BV*) and Delta(15) with
redshift. While the Beta(*BV*) values decrease, consistent with brighter Supernovae Ia (a
Malmquist bias). On the other hand, the Delta(15) values seem to indicate the opposite trend, the
light curves and therefore the supernovae themselves are actually getting *smaller* with increasing
redshift! Why would this be true? Why would we find *smaller* supernova with increasing
redshift? Unless the morphology of supernova are changing, and the spectra indicate that they are
not, we should expect the size of the supernova we actually observe to increase slightly with
distance, a predictable Malmquist type II bias of about 4%.

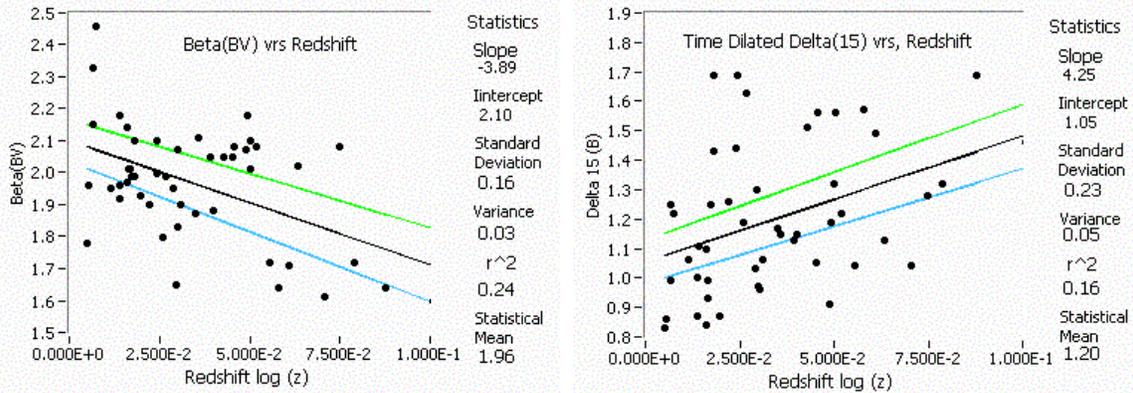

**Figure 6.** (Left) The Beta(*BV*) slope plots a decreasing trend with increasing redshift, consistent with a Malmquist distance bias. (The lines above and below the slope are mean error lines rather than standard deviations.)

**Figure 7.** (Right) The Delta(15) values appear to indicate just the opposite!! Distant supernovae Ia appear to have intrinsically smaller light curves.

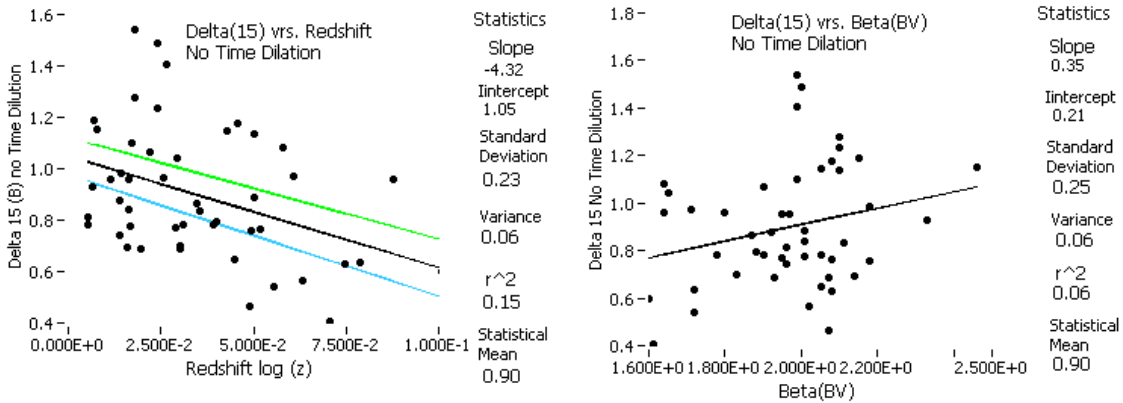

**Figure 8.** (Left) The same Delta(15) vs. Redshift as fig 7, with the time dilation correction removed.

**Figure 9.** (Right) The lack of a strong correlation between corrected Delta(15) and Beta(*BV*) values indicates there is a high degree of independence in these two methods (see **Note 2**).

In Figure 8 the time dilation correction has been removed from the Delta(15) curve. A Malmquist bias is immediately evident: The Delta(15) values now predict increasing supernova light curves. This argument would be virtually irrefutable if a strong correlation existed between the Delta(15) and Beta(*BV*) methods. There is not (Figure 9). *But these data strongly indicate there are at least two independent parameters involve in determining supernovae magnitude and light curve functions, rendering the single 'stretch parameter' approach invalid.*

This comparison uses the *same primary data* cosmologists are using to prove the universe is expanding and by removing the redshift normalization, correction, and stretch factor, and factoring in a reasonable Malmquist bias correction, demonstrate that it is not!

**Supernova Ia Rise Times**
This same argument can be made with the most basic piece of statistical data: Supernovae rise times: In the local universe, the average rise time is 20 days, but in the redshifted universe; it is 17.5 days, which again, would tend to indicate more distant supernova are *smaller* (Li). If this time dilation factor is removed, the high redshift sample has an average rise time of about 25 days. (*Email from Filippenko*). This is *too long* for normal Ia, but not if the distance modulus and the corresponding attenuation factor are underestimated. In this case, the higher redshifted SNe Ia would indeed be over-represented by very high magnitude 'peculiar' Ia, or hypernova. Credence is given to this conjecture by the fact the number of supernova actually found in high redshift surveys represent only a small fraction (~4%) of the expected yield (*Tonry*).

## Very Highly Redshifted Supernova: Type Ia or Type IC Hypernova?

In 1998 a very peculiar supernova was observed. SN 1998bw is the brightest supernova ever seen in the 'local' universe (redshift z = 0.0085). It also has the longest blue light curve, dimming less than one magnitude in more than 28 days (Galama). This is especially unusual because SN 1998sw has a spectrum similar to the core collapse class SN Ic , which normally burn faster and dimmer than supernova Ia. Recently SN 1998bw has been reclassified as a *hypernova* (Iwamoto). The magnitude, light-curve, and associated gamma ray burst may indicate asymmetric burning of a binary pair (Hoeflich). Until SN 1998 was observed, there were no known local supernova events with light curves as long as the small sample of events at high redshift (Leibendgut).

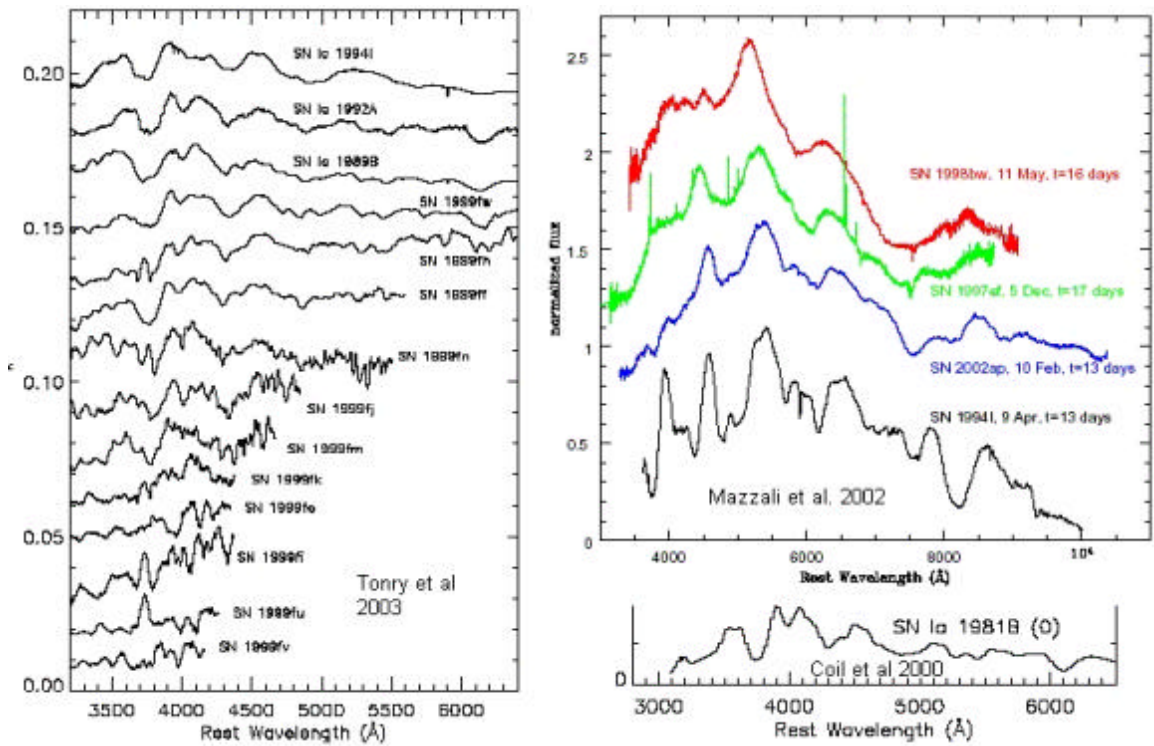

**Figure 10** (*Tonry*), demonstrates how difficult it is to separate SN Ia from SN Ic spectra. The top spectrum is from a local SNe Ic; the next two spectra are from local Ia. Eight of the remaining high redshift spectra have been converted to the rest wavelength from higher redshift, and have been identified as Ia; even though they lack the Silica II dip most characteristic of Ia supernovae. The other three are not identified as to type. The four spectra in *Mazzali* and 1981B (Coil) are from the local SN Ic hypernova, and demonstrate how dramatically the spectrum may vary from star to star. Can anyone really discern SN Ia from SN Ic hypernova at high redshift?

The implications for using supernova Ia as cosmic indicators could not be clearer: If SN Ic can burn in a hypernova state, is it possible for SN Ia to burn as a binary pair as well? Could SN Ic at cosmic distances be mistaken for a binary pair which includes a SN Ia? What separates a SN Ia from SN Ic? In theory, a Type Ia is a critical mass explosion of a white dwarf, while types Ib, Ic and II are core collapse supernovae. Since SN 1998bw, at least four additional Ic have been identified as hypernova (Nomato, see also Bennetti on the difficulty of classifying SN 2002bo). Three of these four are the most energetic supernova events in the local universe (Matheson). If local hypernovae occur, they should also be observed at higher redshift, and since a local SN Ic is the brightest local event ever, very bright Ic hypernovae should also be manifest at higher redshifts, with extremely long time-dilated light-curves. *Where are they*? Are the supernovae observed at high redshifts truly small, time-dilated Ia supernova, or hypernova events like local Ic, with little or no time dilation?

If this hypothesis is true, there should be at least a small number of very faint but *true* SN Ia identifiable at very high redshift. At least one supernova, 2003aj that might fit this description has been found, but it was excluded from the Ia population sample because after correction for time dilation, the light curve is too small! (Riess, 2004)

The *Richardson* magnitude plots imply and the *Wang* multicolor curves confirm a Malmquist bias *should exist* in width of supernova light-curves, yet for all published supernova surveys, magnitude biasing is not considered a factor. If supernova light-curve widths are corrected for magnitude bias, there is no light-curve width budget left for time dilation, and the Wilson hypothesis is not supported. In fact, it fails. (**Note 3**).

## Gamma Ray Energy Distributions and the GKZ limit.

Recent observations demonstrate supernovae explosions also produce gamma rays (Lamb). There should be some resemblance of order between supernovae Ia and gamma ray distributions. Using conventional cosmological interpretations, there is not. When gamma ray photons collide with molecules and electrons in the interstellar medium, they are destroyed, or reradiated at lower frequencies. This limits the distance that gamma rays can penetrate. The theoretical limit for cosmic rays is known as the GKZ cutoff (Stanev). A small percentage of observed gamma rays have known redshifts, which seem to exceed this limit, both in terms of the distance traveled and the power function. (Frail)

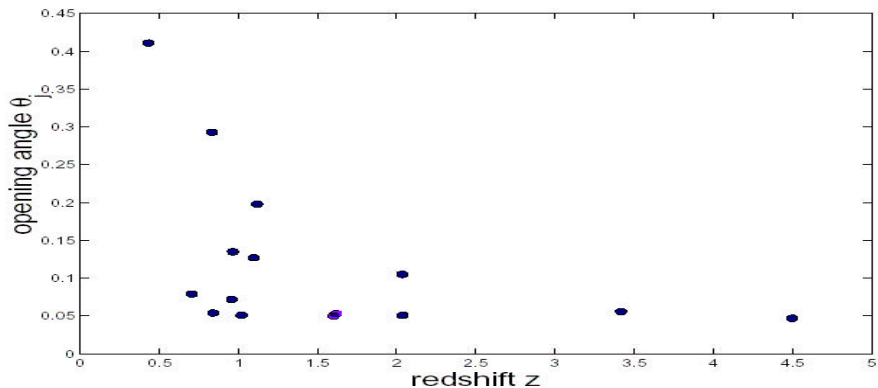

**Figure 11** (van Putten). In order to for gamma rays that appear to originate at high redshift to fit into the same energy distribution as local gamma rays, they must originate as extremely narrow beams with a functional distribution proportional to distance.

The conventional explanation for this paradox is gamma ray beaming (Frail). However, to apply the same power rule to redshifted gamma rays, the apertures of these beaming cones must be proportional to redshift, with the most powerful gamma rays produced by supernovae exploding at the greatest redshift distance. This makes sense if we are observing binary events which focus the energy of a dual detonation (Middleditch). But it also means the more distant events we observe should have longer light curves! Up to twice the length of a singular supernova Ia event.

## Quasar Light Curves: No Evidence of Cosmic Time Dilation

For more than two decades researchers have monitored the light-curves of distant quasars. A statistical data base has been assembled that is baffling: When quasar light-curves are converted to the frequency domain, if time dilation is factored in, the power spectrum of high-redshift quasars is not the same as low-redshift quasars:

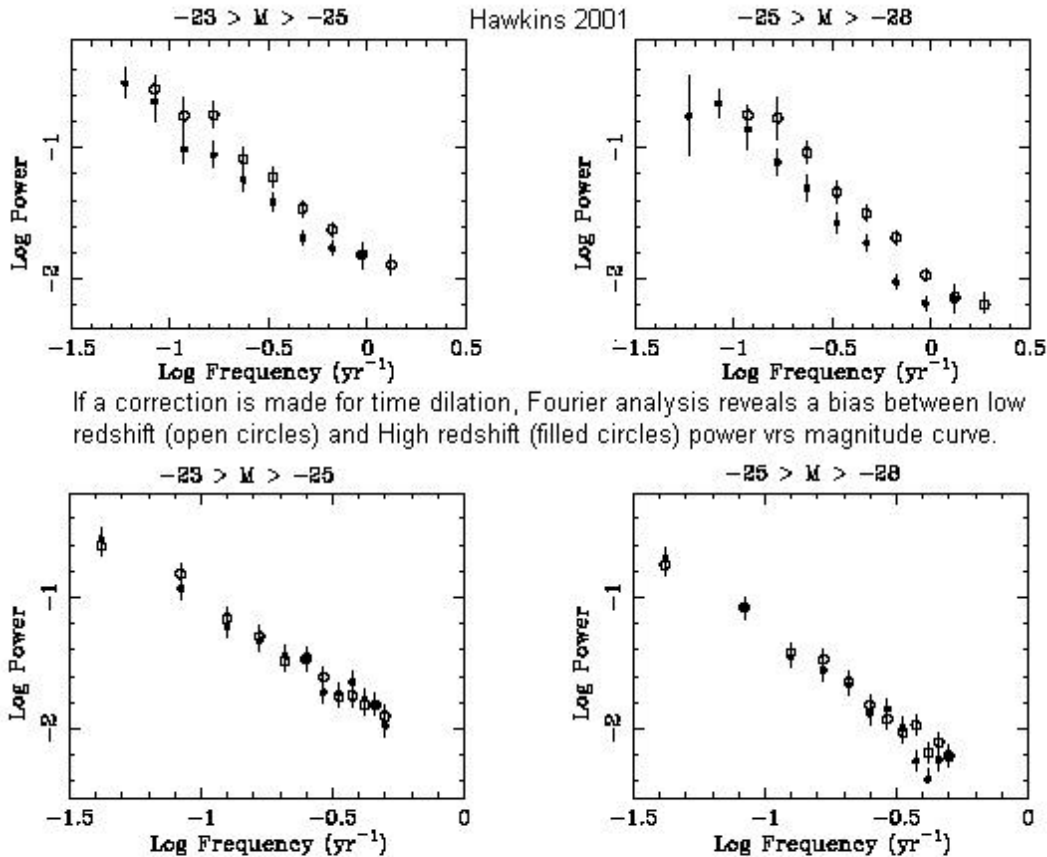

**Figure 12.** (Hawkins) The power functions of quasars degenerate with increasing redshift.

Hawkins notes:

> There would appear to be three possible explanations for the lack of a time dilation effect in quasar light curves, all of which conflict with broad consensus in the astronomical community. Firstly, time dilation might not in fact be a property of the universe…The second possibility is that quasars are not at cosmological distances…The third possibility is that the observed variations are not intrinsic to the quasars, but caused by some intervening process at lower redshift…

The (third) possibility Hawkins cites as the most likely cause of this discrepancy is intervening gravitational microlensing. However, the microlensing solution to this conundrum has been roundly discredited (Zackrisson). Ironically, *this* paper contends all three of Hawkins' suppositions are to some degree correct. Part V of this paper presents additional evidence quasar redshifts are displaced, including proper motion and other periodic effects.

Both the failure of quasar light-curves to demonstrate time dilation, and the relationship between supernovae and gamma rays, raise serious questions about basic SN Ia theory, and challenge the validity of using SNe Ia as analytical tools. The likelihood of a Malmquist bias and the distinct possibility supernova do not satisfy the Wilson hypothesis has not deterred cosmologists from using the magnitude of SNe Ia in the construction of an extended Hubble distance scale, and from it, they have drawn a radical new conclusion: The expansion rate of the universe is increasing.

### Part II: The Effect of Removing Time Dilation from Hubble Magnitude Plots

Hubble magnitude diagrams are important because they indicate, once other biases have been addressed, the correlation between the Hubble constant and the magnitude of distant events: They plot cosmic scaling factors.

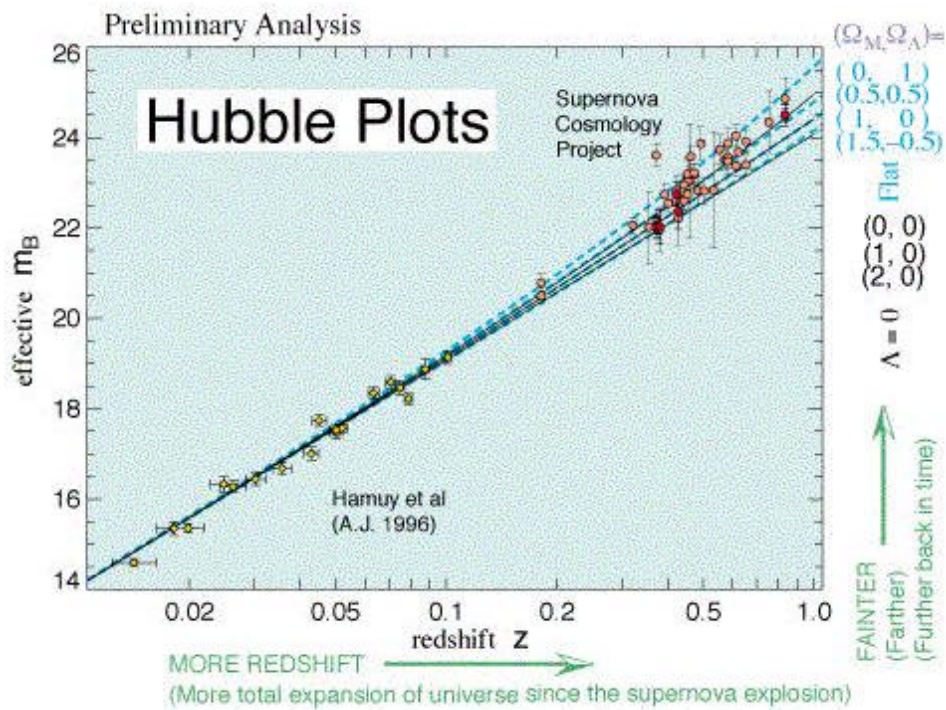

**Figure 13** (SCP). A flat, constantly expanding universe with no dark energy would plot as a straight line on this plot, following the relativistic distance modulus of $\approx 1/(1+z)^4$. The divergence of supernova magnitude towards increasing faintness with distance from this line is considered evidence of dark energy causing an increase in the rate of expansion of the universe.

In Figure 13, the observed magnitudes of SNe Ia follow the predicted relativistic curve for a flat universe quite well, except at high redshifts ($z > 0.2$), where the brightness dwindles faster than the redshift distance predicts. Cosmologists, who were expecting the *opposite* divergence (supernova falling to the lower, *brighter* side of the curve) concluded the rate of expansion of the universe is increasing, rather than decreasing as the Big Bang theory predicted.

However, 'dark energy' induced expansion is not the only conclusion that can be drawn from this stunning departure from theory. *Since the apparent magnitudes of supernovae Ia do not follow the relativistic expansion prediction of the Big Bang, and if the light-curve widths represent a* *magnitude bias rather than time dilation, there is no empirical reason to assume any of the loss of flux is due to expansion.* If the time dilation factor is removed or minimized, it also means the under-luminescence of supernova observed by Perlmutter and others (and attributed to 'dark energy') actually begins at a much lower redshift!

Supernova and quasar light-curves over days, and the apparent magnitude of distant supernova simply do not prove relativistic expansion is the only explanation for all of the redshifting of the universe.

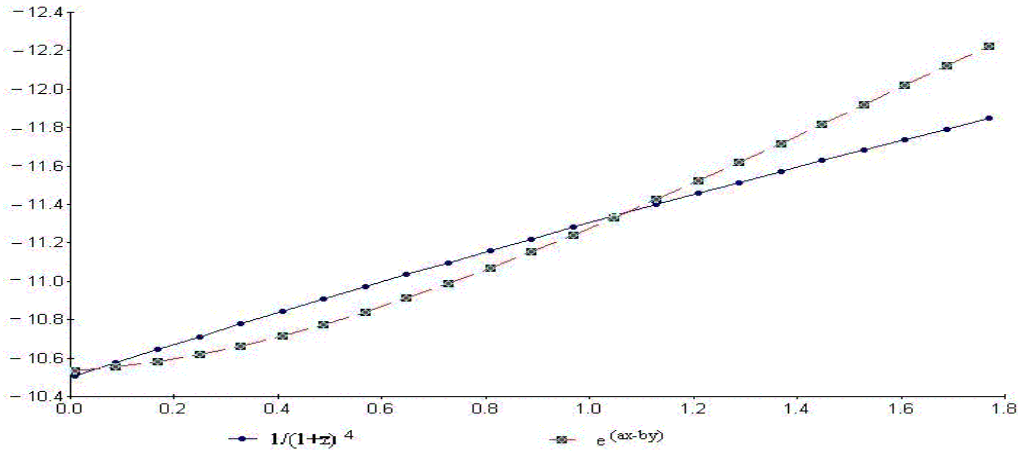

**Figure 14.** A natural log function containing both an increasing and a diminishing exponential term with respect to distance is a better follower of the supernovae Ia light curve attenuation factor and Tolman surface brightness tests than relativistic expansion. This *is* the equation of radiation transfer attenuation of a beam of light between two very distant *bright* objects, $1/(e^{(ax+by)})$.

With no hard evidence of relativist effects in either the magnitude of supernova or in the light-curves of quasars, a new hypothesis is justified. The exponential polynomial function, $1/e^{(ax+by)}$ provides a better curve fit than relativistic expansion to the magnitude/distance function of SNe Ia. If physical processes can be identified that fit a series of exponential functions, an explanation for redshifting which does not require time dilation should exist. This explanation is found in Ramen coherent scattering and in the tensors of polarization of atomic and molecular hydrogen.

## Part III: Coherent Radiation Transfer Functions:  Cosmic Frequency Modulation

In his classical mathematical essays, Chandrasekhar observed:

> A pencil of radiation traversing a medium will be weakened by its interaction with matter. If the specific intensity $I_v$ therefore becomes $I_v + d\ I_v$ after traversing a thickness $ds$ in the direction of its propagation we write $dI_v = -\boldsymbol{k}_v \boldsymbol{r} I_v ds_v$ where $\boldsymbol{r}$ = mass absorption density coefficient for radiation of frequency $v$, *and* $\kappa_v$ is the mass scattering coefficient at wavelength $v$. The radiation transfer function of a given point in space and time can be described in terms of the absorption coefficient by using Kirchoff's law:

$$B_v(T) = \frac{2hv^2}{c^2} \frac{1}{e^{hv/kT} - 1}$$

$\mathrm{j}_v = \kappa_v B(T)_v$      where      $\frac{2hv^2}{c^2} \frac{1}{e^{hv/kT}-1}$ = The Planck Function, where h = Planck's constant and $k$ = Boltsman's constant, , and T is the absolute temperature.

Chandrasekhar's studies of radiation transfer clearly indicate that when a photon is absorbed and reemitted, the emitted frequency will vary according to the kinetic energy of the absorbing particle. In the case of multiple absorptions, the emitted radiation is a heterodyned product of the absorbed frequencies. Since the emitted direction of an incident photon is a vectored sum of the kinetic energy and the incident photon, these functions normally scatter light incoherently and are not considered factors in spectra frequency shifts (**Note 4**)..

Now consider a static universe composed of only two identical galaxies *A* and *B*. Assume the galaxies are 200 Mparsecs apart, with a microwave background and a near uniform column density of molecular hydrogen between them. Also *assume* there is stimulating mechanism for transferring *Planck scale* increments of radiation between the microwave or infrared background and the photonic beams via the molecular hydrogen in the interstellar medium.

Returning to Chandrasekhar's 'pencil of light' analogy, a single cylinder of photons maps the path of light from the source galaxy (*A*) as viewed from target galaxy (*B*). Near the source galaxy, the stimulating photons have a high probability coming from the same direction as light pulses from the same galaxy in a meta-stable absorbed state. At the midpoint between the two galaxies, the photon streams passing each other are of equal size and strength, and there is equal probability that the vectored sum of a radiation transfer will be scattered in virtually any direction. As the pencil of light approaches galaxy (B), the flux in the pencil is maximized from the *opposite direction*. Now most of the stimulated emissions are in the opposite direction. The galaxy (A) flux is scattered at an *exponentially* increasing rate.

Since the incremental shifts are extremely small, this frequency redistribution function should maintain *most* spectral features. Neglecting redshifting, the flux attenuation function may be approximated as $f_o = f_e \Big/ (r^2 e^{rw} e^{w_2/2})$ Where $f_e$ = emitted flux, $w$ is a probability function of temperature, column density, and the radiation transfer rate; $r_2$ = distance from target galaxy $w_2$ is the *increasing* probability of inverse scattering due to the column of light from galaxy B. This effect has been successfully modeled by using many thousands of iterations of the Boltsman relationship:

$$f(x) = x \int_{v=10k}^{12gh} e^{hv/kT_1} dv - x \int_{T=300^\circ K(rnd)v} e^{hv_2/kT_2} dT_2$$

( **see figure 15, Note 5** )

The 300°K temperature used in this equation is justified by the *Eddington temperature* of the local visible light universe. This is an important concept, because in radiation transfer functions, whether a spectral line is *redshifted* or *blueshifted* is a function of the *column temperature*, not the cosmic background temperature. This means the convergence zone between red and blue shifting is occurring in the near IR spectral bandwidths, not in the microwave background range.

While frequencies below the mid-IR range are *Blueshifted*; frequencies above the mid-IR range are *Redshifted.* This produces a dramatic effect in the broad-brush spectral attributes of distant objects: If this expansion were only governed by radiation transfer processes, there would not be both a near IR and a microwave background radiation peak. These two near-blackbody radiation peaks should be merged into one over infinite cosmic distances.

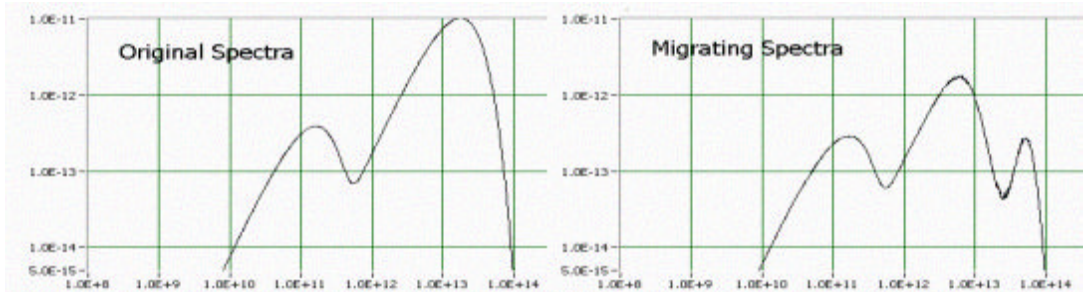

**Figure 15.** Just as radiation transfers cause spectral shifts in distance paths in a prism, the interstellar medium redshifts and attenuates the light we see from distant galaxies. The Spectrum is redshifted, but the primary shape of waveform is unchanged. The waveform 'blue bump' emerging in the migrating spectra is very similar to the spectrum of a very red objects and highly redshifted quasars. This spectrum was produced by subtracting a blackbody with a temperature of $300^{\circ}$K from a $30,000^{\circ}$K blackbody in very small random increments. The $3^{\circ}$K background (first hump) was included in the original spectrum and is multiplied by $10^5$ for display scaling purposes only (**Note 6**).

As can be seen in these examples, radiation transfer functions can duplicate the frequency shifts and attenuation factors seen in the real universe. However, these mechanisms are not simple, and at least three variations of these radiation transfer functions are needed to successfully module the dynamic array of redshifted objects observed in the universe.

At least two of these functions have been formalized in the coherent radiation transfer equations of Moret-Bailly. These functions define radiation transfer in the high flux region in close proximity to quasars, creating the spectral characteristics observed as the Lyman forest.

### Part IV: Coherent Raman Emission of Incoherent Light (CREIL)

Moret-Bailly has used the tensors of polarization and Raman scattering to derive a coherent radiation transfer function for broadband light. In his published works based on laser experiments, he points out "Impulsive Stimulated Raman Scattering" (ISRS) has been studied in laboratories since the 1960's. ISRS effects are observed when differential lasers are trained on tubes containing a gas in the presence of an electric field. In the resulting spectra, the higher frequency is redshifted and the lower frequency is blue shifted.

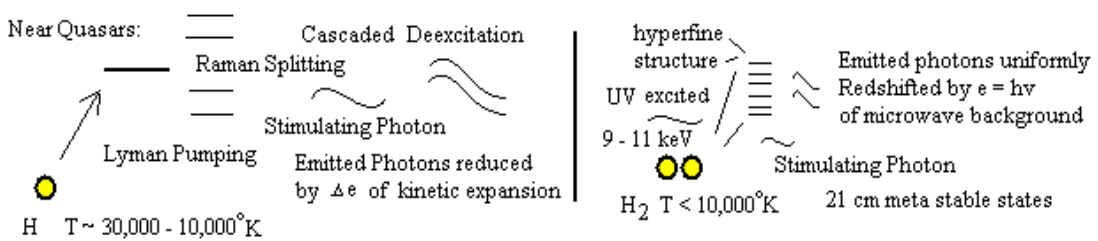

**Figure 16.** In CREIL, the high flux near a quasar or other brilliant source stimulates ionized hydrogen at higher rates than random collisions occur, causing broadband redshifting. At lower temperatures, molecular hydrogen supports a complex CREIL process involving meta stable states.

There are at least two astrophysical environments where stimulated emissions may induce intrinsic redshifts over a broad band of frequencies causing "Coherent Raman Effects on Incoherent Light" (CREIL). The conditions necessary for CREIL exist in close proximity to quasars, and on some scale, throughout intergalactic space.

## The Event Probability of CREIL
Einstein's 1917 analysis of stimulated emissions, in this abridged explanation by Weidner, provides an estimate of the probability of each of these CREIL events:

First assume each atom only has one relevant energy state such that:

$hv = E_1 - E_2$ Using the Boltsman Distribution, $n_1 \propto e^{\frac{-E_1}{kT}}$ and $n_2 \propto e^{\frac{-E_2}{kT}}$

$$\frac{n_1}{n_2} = \frac{e^{\frac{-E_1}{kT}}}{e^{\frac{-E_2}{kT}}} = e^{\frac{(E_2 - E_1)}{kT}} = e^{\frac{hv}{kT}}$$

The number of spontaneous decays per second is proportional to $n$ as $\frac{2_{decay} \rightarrow 1_{state}}{t} = n_2 A_{2 \rightarrow 1}$ Where $A_{2 \rightarrow 1}$ Is the probability for spontaneous decay. The simulated emission and absorption rates may also be written with probability coefficients: $\frac{2_{stimulated} \rightarrow 1_{state}}{t} = n_2 B_{2 \rightarrow 1}$. In thermal equilibrium, the stimulated emission and transmission [ $n_2(B_{1 \rightarrow 2})$ ] rates are the same. The blackbody radiation formula may be used to interpret these *Einstein coefficients*:

$$E(v) = \frac{n_2 A_{2 \rightarrow 1}}{n_1 B_{1 \rightarrow 2} - n_2 B_{2 \rightarrow 1}} = \frac{A_{2 \rightarrow 1}}{B_{2 \rightarrow 1}} \frac{1}{(n_1/n2)(B_{1 \rightarrow 2}/B_{2 \rightarrow 1}) - 1}$$ The Planck blackbody relation is

$$E(v) = \frac{8\pi h v^2}{c^2} \frac{1}{e^{\frac{hv}{kT}} - 1} \therefore \frac{n_2 B_{2 \rightarrow 1} E(v)}{n_2 A_{2 \rightarrow 1}} = \frac{1}{e^{\frac{hv}{kT}} - 1}$$ (Weidner).

At 30,000°K $hv/kT = 0.773$ and the ratio of stimulated to spontaneous emission for visible light (2eV) is approximately $\frac{1}{e^{0.773} - 1} = 0.858$. This ratio is high enough CREIL redshifting events should occur in the extended photosphere of quasars ($T_{Compton} \approx 2 \times 10^7 \, °K$ Saxanov) writing very broad Lyman emission lines. In this type of CREIL, the event probability is proportional to the temperature and the coherent stimulated emission rate drops as the medium cools.

The molecular CREIL process begins between 30,000 and 10,000°K. This is where Lyman absorption lines begin to appear, but since each time an emission peak is encounter the temperature increases, by-stable states occur, stacking, clipping, and truncating emission lines, charting the Lyman forest. As a temperature gradient exists in these zones, there is some spectrum smearing. However, if a strong constant radio signal is present, redshifting is accelerated and spectral lines are preserved. This appears to be why broad absorption line features are found in radio – quiet quasars, while the spectral lines of radio-loud quasars are much better defined. As the RMS temperature cools, the bandwidth of the CREIL process narrows to lower frequencies. As the temperature cools; the CREIL bandwidth narrows and the Lyman forest ends.

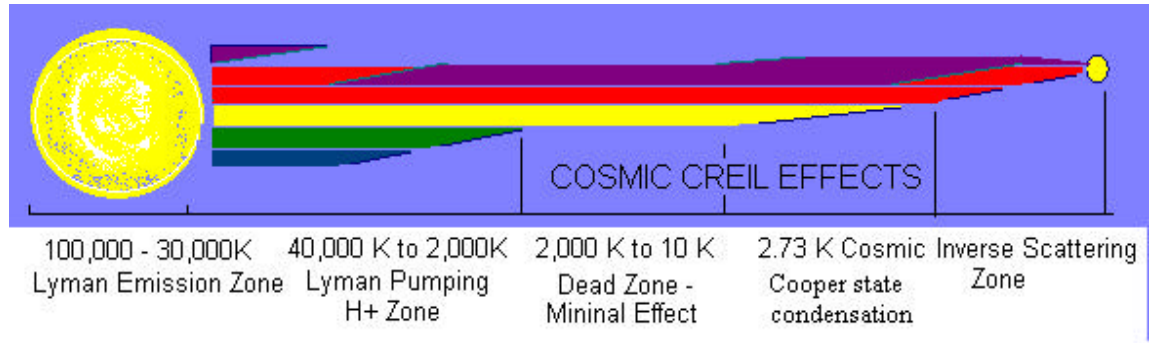

**Figure 17.** As light approaches our galaxy, a photon perched for a nanosecond on a dipole has an increasing probability of being inversely scattered.

We see limited CREIL effects near the sun: (Solar light is actually quite coherent.) The effective high temperature CREIL zone about the sun is relatively small, but the differential redshifting proportional to the radius of the sun results in the band broadening we see in our integrated observations of the sun and other stars: The Wilson-Bappu effect.

Moving into deep space, only the 21 cm nuclear resonance line has the meta-stable states necessary for QREIL processing. The necessary thermal energy is not available for discriminately processing a broadband of radiation through this function. However, the ground state of molecular hydrogen still reveals hyperfine structure about this 21cm line. This is the most dominant wavelength observed by radio astronomers, and provides a rough order of magnitude for intergalactic CREIL: The photon energy at a 21 cm wavelength is 5.9 x$10^{-4}$ eV. At 3°K, Using the same relationship between energy and emissions described above for visible light,

$$\frac{hv}{kT} = \frac{5.9 \times 10^{-4} eV(1.6 \times 10^{-19} J / eV)}{\frac{1.38 \times 10^{-23} J}{°K} 3°K} = 2.29 \times 10^{-2} \therefore \frac{1}{e^{2.29 \times 10^{2}} - 1} = 43.4$$

(Weidner).

Since stimulated emission of the hydrogen ground state exceeds spontaneous events by ~40:1, we should expect to see blue and redshifting in the bandwidths near 3°K. Although this continuance at low frequencies may help explain why the microwave and near-IR background quasi-blackbody peaks do not converge. There may also be other causes, as discussed in part VI.

### Part V: Application of CREIL Processes to the Enigma of Quasars

Nowhere is dark energy cosmology found more deficient than in providing explanations for the quasar population. Much is now known about these brilliant galactic cores (?), but they refuse to properly populate the universal evolution we have so long asserted they track. Writing about the myriads of contradictive studies currently in the literature Jain observes:

> Some of the measured cross-correlations are likely to be affected by the incompleteness of the QSO samples, and most have small samples of galaxies and QSOs with little information on the redshifts of the galaxy population.
>
> Even with these caveats, it is fair to say that there are severe discrepancies between measured QSO-galaxy correlations and theoretical predictions. (Jain 2003)

When the universe is based on CREIL radiation transfers, these conundrums evaporate. First, to summarize and reiterate CREIL concepts:

- 1) Redshifting is a complex combination of intrinsic, proximal and cosmic Doppler and non-Doppler effects.
- 2) In high flux, high temperature regions, the redshift rate per unit of linear distance is a function of net flux vectors, temperature, pressure, spectral energy, relative speed, and electromagnetic & gravitational field strength.
- 3) Sources of high flux scatter light both approaching and receding from the source.
- 4) A correct reading of the distance modulus differentially displaces quasars and other very blue objects.

The following example from *The Host Galaxies of Radio-Loud and Radio-Quiet Quasars* illustrates the paradigous shift necessary to interpret these translations. Dunlop notes:

> The effective study of quasar hosts at high redshift is still in its infancy. However, already it is becoming clear that the mass offset between RQQ and RLQ hosts [described in paper] appears to grow with increasing redshift lending additional credence to its reality. Specifically, for the same nuclear luminosity, RQQ hosts at z ~ 2 appear to be a factor of 2-3 less massive than either their low-z counterparts or their z ~ 2 radio-loud counterparts.

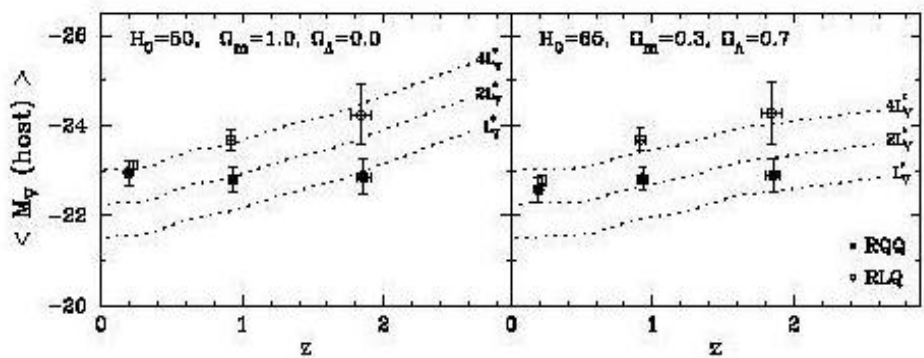

**Figure 18** (Dunlop). Radio-loud quasars (RLQ, open circles) *appear* to increase in magnitude with increasing redshift, but while radio-quiet quasars (RQQ, filled circles) maintain the same magnitude trend, the host galaxies *seem to* become smaller with distance (see quoted text). These two graphs differ by the standard (Peebles) and increasing (Goldhaber) distance modulus. The dotted lines plot the "evolution" of L, 2L and 4L elliptical galaxies into the past, also an artifact of the distance modulus.

Why would radio-quite quasars reside in smaller galaxies in the past, while radio-loud quasars were both brighter and lived in larger galaxies? The electromagnetic field coupled to Radio-loud galaxies *predictably* and dramatically increases the CREIL redshifting rate! Radio loud galaxies, at a redshift of two, are much closer and therefore appear to be brighter and larger than the distance modulus predicts. On the other hand, radio quiet galaxies, just like supernova, are only slightly closer than the redshift predicted distance modules, they appear just as luminescent, but since the distance modulus underestimates attenuation factors, mean galaxy sizes are overestimated.

This is why some surveys disproportionately place quasars in near low redshift clusters, while other surveys indicate an anti-correlation (Jain 2003). Radio-loud quasar types are distance miss-casts, radio quiet families, less so. Since radio active galaxies imply intrinsic redshifting, active galaxies which reside in closer proximity than the redshift indicates, should statistically appear to have higher absolute magnitudes than the general population. This is also true (Barton).

**Evidence of CREIL 1: The Paradoxical Lyman Forest**

The Lyman Forest was first theorized by Gunn & Peterson, Scheuer and Shklovski in 1965. In theory, the intergalactic space is occupied by neutral hydrogen commoving with the expansion of the universe. As light from distance sources passes through these clouds, the Lyman UV peak (@1215.67 Å) absorbs and attenuates these light sources. As the relative motion of the intervening clouds of hydrogen decrease, the Lyman absorption lines move through this spectrum, creating a 'picket fence' of lines greatly attenuating the spectrum between the rest frame of the redshifted source and our telescopes.

The Lyman forest is assumed to exist because the spectral attenuation of many quasars is 'nearly' as predicted. However, there are at least three problems: First in virtually all quasars, there is NO observed forest within a redshift of about z=0.5 to the source. After this break, there is a 'continuous forest', with few if any broken lines. The second problem is that Hubble data indicates that of quasars less than a redshift of about z=1, "a constant commoving density of objects" best describes Lyman events while at redshifts above z = 1 the "Ly cloud column densities $N$ are distributed according to a power law " (Raunch). Why should cloud density be constant at low redshifts, but follow a power law function at increasing redshift, then disappear completely for the last 0.5 redshift in front of every quasar? A third problem is that quasar spectra contain a 'blue bulge', not handled well by any dark matter theory. (Kishimoto, Murphy, Reichards)

CREIL provides a drastically different explanation for the nature of quasar spectra, placing most of the spectral shift in close proximity to the quasar. As the photosphere of a quasar cools, the initial spectra, consisting primarily of the Alpha, Beta and Gamma hydrogen lines are written into the spectra. During the high flux CREIL redshifting near the quasar, virtually all of the hydrogen is ionized and there is little absorption. As the flux is diluted, the 'hot CREIL' effect diminishes. Once the gas temperature has cooled to a near uniform temperature, the probability of 'cold' CREIL increases. During cold CREIL shifting, the ratcheting effects of the absorption peaks in molecular hydrogen clip and attenuate the redshifting lines as they enter the 'forest' (Jensen). This is why the Lyman forest AWAYS begins at a red shift of about 0.5 from the redshift of the quasar. Astro-physicists have always been at a loss to explain this behaviour in terms of hydrogen clouds.

There is highly collaborative evidence emerging from XMM-Newton and Chandra studies of quasar X-ray spectra. XMM-Newton has resolved two sets of fluctuating broad line spectra emerging from a single quasar, both at highly relativistic speeds (Chartas). The classic explanation involving massive quasar winds, will never model this behavior with Newtonian kinetic. However an electromagnetic quadrupole could easily impose differential CREIL effects which can predict and model this clearly non-Machian behavior.

**Evidence of CREIL 2: Periodic Quasar Abundances: An Artifact of CREIL Redshifts**

A periodic feature in the redshifts of quasars has been identified by Arp and characterized by Bell, Russell and others (Bell & Comeau). There is an intrinsic redshift in quasars at fixed multiples of 0.062 and other harmonics (Arp, Russell). These periodic functions have been well defined but are extremely difficult to explain in an expansion and evolutionary context: Where could a periodic function have originated? Why would such a harmonic function survive the chaotic clumping of galaxies through eons of expansion?

This periodic function is explained in terms of the coincidence of Lyman impressions in the CREIL. By arranging the quasar quantum numbers supplied by Bell, Moret-Bailly[2] has identified a meaningful relationship between these quantum numbers and Lyman α and β lines that creates periodic increases in spectral intensity (**Figure 19**). Quasars at periodic multiples appear brighter to us because of the strobing effects of Lyman alpha, beta and gamma absorptions. The ability of CREIL to resolve this cosmic periodic function, which has no valid function in a uniformly expanding universe, has significant merit.

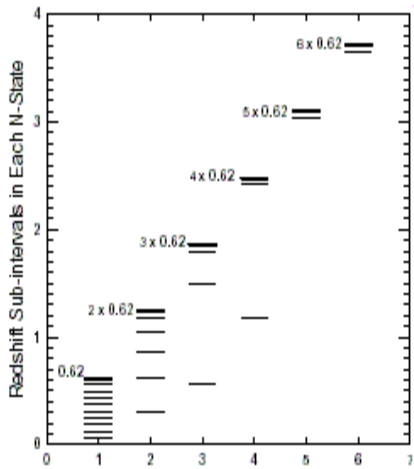

Coincidence of
Rydberg Approximations of $H_{\alpha\beta\gamma}$ Lines

$$z_{(\beta\ resp.\ \gamma),\alpha} = \frac{\nu_{(\beta\ resp.\ \gamma)} - \nu_\alpha}{\nu_\alpha} \approx$$

$$\frac{1 - 1/(3^2\ resp.\ 4^2) - (1 - 1/2^2)}{1 - 1/2^2}$$

$$z_{(\beta,\alpha)} \approx 5/27 \approx 0.1852 \approx 3*0.0617$$

$$z_{(\gamma,\alpha)} = 1/4 = 0.25 = 4*0.0625.$$

$$z_{(\gamma,\beta)} \approx 7/108 \approx 0.065.$$

**Figure 19** (Bell & Comeau). The harmonic magnitude and concentration peaks identified in the quasar population. The Rydberg approximations are taken directly from Moret-Bailey[3].

### Evidence of CREIL 3: Proper Motion in Quasars?

Using a multi-decade survey, McMillian has observed proper motions in quasars of over 50μas/yr. This places quasars squarely in the local universe. If this data is collaborated, intrinsic redshifting in (or near) quasars is a fact. Since quasar galaxies are at the same measured redshift as quasars, a CRIEL-like process is necessary to produce this *sympathetic redshifting*. The conventional explanation for this 'apparent' proper motion, microlensing by intervening objects, is being severely challenged by the obstinacies of these proper motion trends (Oguri).

### Evidence of CREIL 4: Unmasking the Tully-Fisher Relationship

The Tully-Fisher relationship is based on a predictable correlation between the rotational velocity of the disk and the magnitude of the galaxy as a whole: The rotational velocity is estimated by observing the differential redshift across the plane of the galaxy, either directly or imply by H or other line widths in the spirals of the galaxies: The faster the rotation, the more luminescent the galaxy. There should be some degree of correlation, (since faster radial velocities imply greater rotational velocities and therefore greater mass), however the relationship holds over a very wide range of magnitudes and distances, implying there is a transcending fundamental relationship between the rotating disk and central galactic mass. Luc Simard Observes:

> The Tully-Fisher relation highlights two remarkable properties of galaxies. First, the mass-to-light ratio of galaxies must remain constant over a broad range in luminosity since the relation remains tight over at least 7 magnitudes (a factor of 600 in luminosity!)...Second, there must be a conspiracy between the disks (luminous matter) and halos (dark matter) of galaxies such that both components ``know" exactly how much each one should contribute to the mass within a characteristic optical radius as a function of total galaxy mass.

The CREIL effect, the impinging of the flux from the core of the galaxy on the disc of will cause a broadening of the H lines used to estimate the rotational velocity of the disk, but since this involves a stimulated emission mechanism, the existing spectral distribution will be enhanced, in both the receding and the advancing sections of the disk: The brighter the core, the greater the line broadening and the more stimulated emissions in the disk. As this flux is passed through the successive bands, the higher frequencies of the inner bands are imbedded in the flux passed from the core to the fringe of the disk. This increases the apparent rotational speed of the disk.

The robust nature of the Tully-Fisher relationship *may* be a product of CREIL. Since more of the central flux should be viewed approaching us than receding from us, a bias should exist between the 'red shifted' and blue 'shifted' sides of a galaxy. Since the flux from the bright galactic center should fractionally increase the H line wide with increasing distance to the observer, there should be a slight degeneracy in the Tully-Fisher function, evidenced by an increasingly non-random distribution in the spatial orientation of galaxies with increasing distance. This degeneracy in fact exists, and is interpreted as an evolutionary morphological change in the Tully-Fisher function with increasing distance. (**Note 7**)

**Evidence of CREIL 5: Very Red Objects: Resistive Reddening and Relative Redshift**
Surveys have indicated there is a statistical over abundance of very red objects (VROs) in close proximity to quasars (Hall, Wold). Why would this happen? According to CREIL theory, quasars are displaced in space making the redshifted galaxies are more distant. As the flux from these galaxies approaches the quasar, the high UV flux from the quasar enhances the CREIL interaction in the *opposite* direction. This attenuation or *resistive reddening* is essentially Rayleigh scattering, and since the thermal energy is also elevated above background levels, the radiation transfer appears very similar to dust reddening. There is also a reported overabundance of red galaxies near quasars (Best 2003), which is also an artifact of this effect.

**Evidence of CREIL 6: Ultra-Luminescent Infrared Galaxies**
Two earmarks of ultra-luminescent infrared galaxies are high levels of molecular hydrogen and the appearance of a dusty environment (Sanders). Many of these galaxies, such as ARP 220, appear to involve collisions of galactic systems. The huge clouds of hydrogen and the bursts of flux from these active centers scream of stimulated emission reddening and redshifting. These objects often produce "humped" spectra similar to **figure 15**. (Similar effects are also observed in hydrogen rich dusty environments near local stars. These features also suggest molecular and atomic hydrogen may differentially amplify light at resonant wavelengths (An). Galaxies that contain quasars in there centers are often both severely reddened, and redshifted relative to other galaxies at the same apparent distance. This is why quasar galaxies appear to be so huge. They are average sized galaxies, as their morphologies tend to indicate (Boller).

**Evidence of CREIL 7: Sympathetic Excitation: The Cepheid and Tully-Fisher Distance Scale Bias and Correlation of this Variance with nearby Galaxy Brightness**
A surprising correlation has been identified between the magnitude of some galaxies in tight clusters and otherwise indefinable bias between Tully-Fisher and Cepheids distance scales. By factoring in the 'flux impinged' on a galactic system by nearby galaxies, an error correction factor can be made for Tully-Fisher galaxy distances, which eliminates the variance between Tully-Fisher and Cepheid distances. *Hodge and Castelaz* use a family of correction factors, depending on the galactic alignment, but as the distance increases, these factors converge to a single linear regression.

It seems ludicrous to suggest the flux from nearby galaxies significantly affects the magnitude of an entire galactic plan. However, the converging light from each galaxy increases the probability of a stimulated emission in the general direction of our galaxy, and also increases line bandwidths. Although at first glance this may appear contrary to quasar reddening effects. In a quasar, a higher percentage of the quasar photons are high-energy particles, these causing a localized heating of the microwave background in much the same way as cosmic rays, randomizing the directional scattering, reddening the objects spatially located behind the quasar.

## Evidence of CREIL 8: Ca II & Mg II Emission lines

In 1957, Wilson and Bappu discovered the width of the CaII K emission line is proportional to the absolute magnitude of certain classes of stars and can therefore be used to calculate distance. A similar method has been developed using the Mg II K emission line at 2796.34. Although a high degree of correlation exists between these methods across a fairly high range of luminosities (CC=0.9), it breaks down for very high flux stars, and breaks down completely in binary systems. Cardini et al observes:

> We observe that the corresponding correlation coefficient is fairly high in spite of the ~ 6 order of magnitude span in line luminosity. This is a remarkable result in itself, which should deserve proper study for its implications in the understanding of line broadening mechanisms in the chromospheres of stars…
>
> The distribution of [binary] stars in the [Mg II K peak width *vs.* magnitude] does not show any regular pattern…On the contrary, [binary] stars are closely correlated in [Mg II K peak magnitude *vs.* magnitude$_V$ ]…Such a strong correlation is also found for normal stars…but [normal stars] are systematically fainter in the MgII K line by a factor of ~ 10 compared to [binary] stars.

Just as with the Tully-Fisher relationship, the width and magnitude of these peaks and the confusing correlations are fairly simple to model using a CREIL process: The bigger and therefore brighter the star, the more redshifting that occurs near the radius of the chromosphere, broadening the peaks. When two stars are close to each other, the intense flux from each star creates an intrinsic stimulated redshift and scattering throughout the elementally enriched space between the pair, greatly increasing both the width and the magnitude of the metal emission lines.

## Evidence of CREIL 9: Optical Polarization: Complimentary Redshifting

Evidence of this broadband CREIL mechanism is found in the correlation of the polarization of light from quasars that are close to each other. This polarization is directly proportional to both the quasar proximity to each other and distance from the observer. It also appears to be independent of the intervening medium (Jain, Pankaj). Polarization is a natural consequence of stimulated emissions (Lyutikov). This implies flux from nearby quasars contaminate each other by hijacking the adjacent spectra during the catalyzed transfer, consolidating details in the images of two quasars in close proximity to each other relative to the distance the light has traveled.

## Evidence of CREIL 10: Arp's Quixotic Visual and Radio Images

In papers too numerous to properly cite, Arp and others have argued for decades the coincidence of Quasars with local galactic clusters is beyond a mathematical absurdity (Lopez-Corredoira). That astronomers continue to doubt these convincing visual and tomographic radio images is one of the perplexing consequences of the power of preconceptions on the interpretive functions of the human mind.

## Part VI: Removing Evolutionary Artifacts from Cosmology

There are many reasons theoreticians have found it easy to dismiss steady state or "tired light" models. It has already been demonstrated how an exponential distance modulus can approximate the $(1/1+z^4)$ relativistic attenuation factor, but other problems must be addressed. Additional arguments in favor of an expanding universe include:

1) Steady state models cannot explain evolutionary trends without violating the Cosmological Principle.
2) There is no solution for Obler's paradox: If the universe is infinite, the sky should be white, or at least very red.
3) There is no causality for the Gaussian cosmic microwave or infrared backgrounds.
4) Static models provide no testable explanations for light element abundance and cosmic synthesis.

Any attempt to displace the Big Bang's hold on cosmology must address these issues, all of which have been answered successfully by the standard model. Each of these other items will now be addressed.

### The Era of Quasars, Early and Late Galaxies and other 'Evolutionary' Trends

The "Copernicus Principle" states that any trend we observe which makes the earth, the sun, or even our galaxy *appear* to be singularly unique in the entire universe must have an alternative explanation. One of the subtle problems of the Big Bang theory is that the current era *does appear* to be unique: Major evolutionary changes appear to have occurring as peak events in the recent past. For example, there is a peak in both the magnitude and count of quasars at a redshift of ~z=2 (Wolf, Floyd). And there is a similar peak in blue/late type galaxies at a redshift of ~0.05:

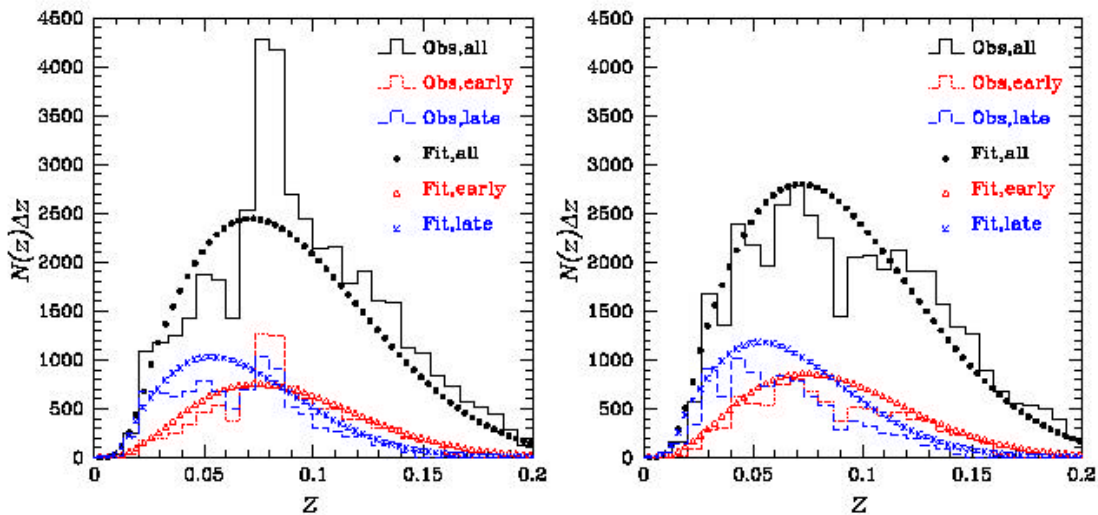

**Figure 20** (Hikage). From an early release of Sloan Galaxy survey, comparing distribution of early and late type galaxies. This distribution is based on the relativistic distance modulus. This trend is an artifact of both the incorrect distance modulus and intrinsic redshifting in the brightest galaxies disproportionately skewing the redshift. Compare with figure 21. (The distributed count of 'all galaxies' is much greater because not all galaxies have been identified as 'early' or 'late'.)

But while these evolutionary trends pose modeling difficulties for Big Bang theorists, these trends appear to be impossible to explain in a steady-state model, where *no* evolution is allowed.

However, non-evolutionary models can be developed, using the fundamental rules of CREIL redshifting mechanisms *and* the resulting exponential distance modulus:

1) Very bright, blue objects, such as QSO, starburst galaxies, and blue late-type galaxies are intrinsically redshifted.
2) This intrinsic redshifting displaces these objects relative to red galaxies.
3) The rate of attenuation of all objects increases with increasing distance.

If these postulates are correct: the following trends should exist:

1) There should *appear* to be no quasars or very blue objects in our galaxy, and few if any our local galaxy cluster.
2) Since these Blue objects are 'distance displaced' by intrinsic redshifting, they should not statistically appear to be in the centers of clusters, co-mingling with red galaxies.
3) As redshift distances increase, the probability of finding blue galaxies in the centers of red clusters should increase (because the absolute distance becomes less well defined.)
4) A mathematical compression should occur in the blue object count, peaking at a redshift equal to about twice the estimated intrinsic redshift.
5) High redshift supernova detection rates should fall short of expectations.

All of these statements are true! No late-type blue galaxies appear to be in our immediate vicinity, nor are there any local quasars. Blue galaxy populations first appear to increase, then decrease with distance, they avoid the centers of clusters in local space, but are increasingly found in galactic cluster cores with increasing distance: the Butcher-Oemler effect (Goto). Note that if all of these conditions were not met, this 'differential redshift' model would not work. But it does: All of these 'evolutionary' trends are artifacts of CREIL-type radiation transfer functions.

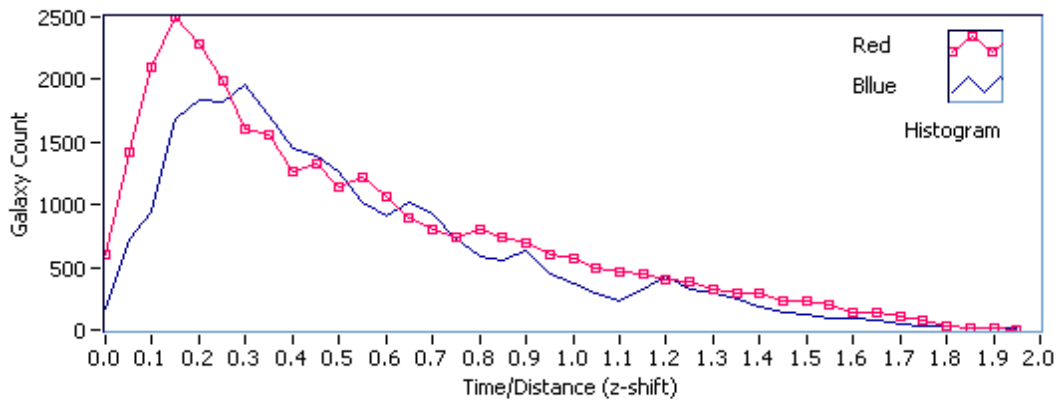

**Figure 21.** A binned Monte Carlo simulation using Arp's sequence for discrete intrinsic redshifts of blue galaxy clusters, (0.06, 0.3, 0.6, 0.91...), these increments are randomly added to the blue population only. This causes the blue population to be underestimate at z = 0, overestimated at z=0.3, and fluctuate periodically at increasing distances. These apparent periodic effects are also present in quasars surveys.

This simple model demonstrates how the periodic concentration of 'blue galaxy clusters' observed by Arp is caused by intrinsic redshifting of the 'blue' galaxies at discrete intervals, displacing and compressing the local count at z = 0.3. If this effect is combined with an exponential attenuation factor that increases with distance, a powerful argument can be made that all galactic evolution is indeed an artifact of our lack of understanding of the nature of radiation transfers in intergalactic space.

**Obler's Paradox**
This analysis offers three solutions to Obler's paradox, all of which should contribute to the darkness of the sky: 1) When is a transparent object black? When light source is totally internally reflected. In portions of the sky where the flux source is most distant, the probability that the light approaching us has been inversely scattered approaches infinity. 2) The sky is dark because it isn't: For transmitted light, the radiation transfer functions describe herein funnel both higher and lower frequencies into the infrared continuum: The cosmic infrared background has a much higher energy density than any big bang theorists predicted, in fact it may have more energy than the optical background (Elbez). 3) Cosmic light sinks: If black holes, or some permutation of these accretion rings exist, a fraction of the radiant energy disappears into these horizon events. (An additional solution to Obler's paradox is found in part VII.)

**Einstein Coefficients in the Cosmic Microwave Background**
The current cosmology developed about the cosmic microwave background (CMB) provides us with sophisticated interpretation of the fine microwave spectral detail. The problem is, this structure does not support the Big Bang Theory. The quadrupole moment is less than expected (Cline). The "Cold Dark Matter Peak" misses *a prior* attempts to model cold dark matter by a country mile. In fact, the WMAP secondary peak is exactly where MaGaugh predicted it should be for a NO cold dark matter model (McGaugh). What little structure there is, outside of galactic pollution, correlates very well with known x-ray sources, so much so it is impossible to separate large structure Sachs-Wolfe effects from x-ray Compton scattering (Boughn 2003). The CMB failed to meet the predictions of Sachs & Wolfe, failed to satisfy the refinements of Peebles & Yu, it failed to satisfy the angular resolution calculated by Wilson and Silk, and it even failed to meet the most modest power spectrum requirements of the elusive cold dark matter (Wright 2003). It is these and other failures of the big bang that have burdened us with 'dark energy': A non-solution to the cosmos Peebles has dubbed "Embarrassing" (Peebles 2002). The thermal background is the muddiest part of the universal energy pool, contaminated by galactic, cosmic and solar sources. Divining precision cosmic parameters from this sinkhole is witchcraft.

The primary cosmic microwave background can be expressed as a single black body function. This is consistent with the radiation signature Einstein coefficients predict in laser stimulated emission events (Prigara). Like laser light, CREIL is a stimulated emission process and derives the same Einstein coefficients as a laser process. In the local universe, the rms temperature is 300K. CREIL transfers energy from this infrared continuum to the microwave background. This is a necessary signature of steady-state universe in a completely relaxed equilibrium.

**Light Element Abundance**
This implied quasi-steady state universe requires an explanation for the copious quantities of hydrogen observed in the universe. However, any theory discounting continuous light element synthesis, such as the big bang, has a difficult time explaining why 'old' systems contain young star forming in areas depleted of heavy elements; and why there are areas of heavy element abundance now apparent in many systems at very high redshifts. Lilly, Carollo and Stockton concluded after shifting through three redshifts of data that there is "only a modest change in mean metallicity compared with the present." (Even this apparent metal depletion is in part a function of the lose of fine spectral detail during CREIL redshifting.) They also conclude "faint blue galaxies", once thought to be precursors to own cluster of galaxies, contain too much metal to be on the same evolutionary track as our own galaxy (Yoshiaki).

Areas of new star formation, high in light elements that have been identified in very old clusters provide a tantalizing hint of a rejuvenation process in the heart of active galaxies. The engine required to complete this hydrogen synthesis may or may not violated the second law of thermal-dynamics. If the universe is gradually winding down, there had to be a creation. If point violations of the second law are allowed, matter exists simply because it has always existed.

The radio source count is also much higher than predicted by the big bang without introducing new parameters (Wright WebPages). This too, is a signature of an infinite redshifted, but not expanding universe. Galactic clusters may be collapsing, but the universe as a whole exists in a vast webby equilibrium, consistent with the Cosmological Principle. (**Note 7.**)

## Part VII: The Friability of Space

The preservation of spectral features and consistency of the redshifting across eons requires a very broadband redshift function. This *speculative* section explores a mechanism that may also be responsible for light element synthesis, MOND phrenology, and a coupling between gravitational and electromagnetic field forces.

Take a single molecule of ozone in the outermost atmosphere of our planet: It's position, motion, and path though space is dictated by the net sum of all the electromagnetic and gravitational forces between the two planets. The position of the molecule could also be effected by a photon, splitting the molecule. These changes in mass and momentum will affect the orbits of both planets and do so at the speed of light. The simple fact is every particle of matter must be capable of transmitting and receiving an inertial or gravitational force vector from virtually anywhere. To deal with the tensors necessary to transfer this energy, Bohr added an infinite 'zero point' field as a pool for gravitational energy reserves. A unified theory that could eliminate this perplexingly endless energy pool has the potential to resolve other cosmic incongruencies, such as dark matter or MOND effects, Obler's paradox and, in the absences of a Big Bang, light element synthesis.

**Bose-Einstein Condensation - a Celestial Radiation and Gravimetric Transfer Mechanism**
Part of the fallout from the displacement of quasars by the CREIL redshifting process is that the interstellar molecular hydrogen content is much greater than current estimates. This is because these estimates are based upon Lyman absorbencies in the path of quasar light (Prochaska). This puts the mean molecular hydrogen density at near galactic levels throughout the observed universe: ~0.15/cc^2, (Izmodenov).

The tensors of polarization clearly indicate a coupling exists between photons and gas molecules (Derouich). To couple photons with discrete interstellar molecular hydrogen in deep space, a new application of a known electromagnetic transfer mechanism is proposed: A discrete low temperature collapse of molecular hydrogen into a boson via a Bose-Einstein condensation could facilitate this energy transfer function. That this relaxed Cooper state exists, is evidenced by experiments demonstrating both superfluidity and conductivity in molecular hydrogen (Grebenev).

Since at low temperatures hydrogen normally condenses into a solid, coaxing superfluidity from a hydrogen molecule is a technically complex, almost wizical process. But in interstellar space, discrete molecules may condense into Cooper states at temperatures near the microwave background. The properties of this very unique isolated quantum state are difficult to predict, but the energy transfer function is governed by the Bose-Einstein distribution:

$$f_e = 1 \Big/ (Ae^{E/kT} - 1)$$

For photons, $A = 1$. This is critical, because it allows the absorption and transmission of virtually *any* photon, regardless of frequency. Substitution of this function into Chandrasekhar's radiation transfer equations produces nearly identical results in the superhetrodyned process modeled in Part III.

Now place a molecule of hydrogen 10 AU from the sun, but in a direct path between the earth and the Pioneer 10 space probe. The earth station is broadcasting an 11cm radio wave toward the Pioneer craft. The Cooper effect converts the molecule into a boson, allowing this molecule to absorb this photon into a meta-stable 1S orbital. When a second absorption results in a stimulated emission, the rules of radiation transfer require a net thermal redistribution, of ~h $(v_1 - v_2)/k$ T from the microwave background to the emitted photons, where $v_1$ is the frequency of the 11 cm radio waves and $v_2$ is the frequency of the 21 cm spin state. This function operates on both the radio waves broadcast to the Pioneer and the returning frequency, which is governed by a phased locked loop in the Pioneer probe. The resulting net *blue shift* of the 11 cm photons is negligible, except when measured as a cumulative effect at very great distances. This frequency shift is interpreted as a Doppler effect, inexplicably accelerating the pioneer probe toward the sun (Anderson). It is almost chilling to note that this acceleration rate is very close to the current Hubble constant for the universe. If this description of the radiation transfer function is correct, this *is* the Hubble constant.

There is a second hypothetical transfer mechanism, which may be piggy-backed into this same Cooper effect transition: Absorption into the 1S orbital couples the photon to the root 'gravitational' force of the molecular nucleus. To escape from this orbital, a gravitational transfer is also necessary. Consider Peebles' speculation concerning gravity and Dirac's observations:

> ...The ratio of the present Hubble length to the classical electron radius is close to another enormous number, the ratio of the electric and gravitational forces between a proton and electron.

Using this Cooper state coupling, the cosmic redshifting becomes the friability factor for the entire universe, coupling gravitational forces with the radiation distance traveled! Gravity is no longer a separate component, but the net vectored force function of radiative fields. This electromagnetic mechanism would provide the inertial and gravitational coupling so long sought by Einstein: The same quantum field effects responsible for superfluidity and superconductivity supply the net vectored field strength quantified by Newton as the gravitational and inertial forces. There are several reasons for speculating there is a gravitational component in radiation transfer functions:

1) The Sloan Digital survey makes it clear thready structures permeate the known universe, But there are also areas of space which appear to be void of *all matter*. This coupling mechanism implies mass is required to transfer all radiation functions, allowing voids.

2) For an infinite universe based on radiation transfer functions, Obler's paradox requires a convergence of the microwave background with the infrared continuum. If the molecular hydrogen concentration is uniform throughout all space and time, this should be the case. However, if *true material voids* exist, and the Sloan digital survey indicates they do, the net vectoring of radiation transfer functions routes photons around these areas. In this case, the continuum in the near-IR and the microwave background never converge. However, the near-IR spectra should and does indicate an increasing power function with distance, which is at odds with the available light budget for dust extinction (Zubko).

3) Both MOND phrenology and dark matter theories fail to predict observed celestial mechanics in certain situations. If gravity is truly an electromechanical force, coupled through the mechanism described above, the natural acceleration of objects at the fringe of a galaxy follow the curvature of the galaxy as the net field strength subsides into the near voids. Photons and the 'gravitational' component of radiation follow the only available tensors along this galactic fringe. This clearly redefines inertial energy, allowing essentially nonlinear paths of least resistance when there is no supporting material tensor.

4) The very concept of friability and the management of the tensor fields needed to balance these forces limit the abject distance a radiation function may exist in an infinite universe. This is apparent in the net exchange of energy required to predict constantly changing gravitational paths for the tiniest particles. In essence this adds a time/distance component to basic equations governing mass distributions (Derouich).

5) Tolman-Bayin dynamics can be used to localize gravitational effects and eliminating the need for either a big bang or a cosmological constant to prevent the total collapse of the universe (Ray). They also eliminate the singularity created in a black hole collapse. The Chandra x-ray schematics of the crab nebula demonstrate how the remnant of a supernova whirls into a cyclotronic accelerator, which on a grand scale may be capable of reducing complex matter into a primal state. (We call mature forms of these light element regenerators galaxies.) Massive galaxies that are ejecting massive quasars like a Roman cannon, as observed by Arp, are in complete harmony with this concept.

Finally, assigning gravity to an electromagnetic tensor, a very weak electronegative force, provides a solution to inertial coupling which has never been realized. Discrete integration of the field strength of any electron orbital configuration where the electron radius, $r_e \gg r_p$ always yields a net attractive force. Could gravity be this simple?

## Summary

This thesis has demonstrated there is a workable solution to the cosmos well outside of current theoretical constraints. The light curves of supernova Ia and quasars do not statistically track time dilation. The apparent magnitudes of SNe Ia do not follow the relativistic expansion curves. A system of exponential equations can be used to plot the magnitude of SNe Ia. Radiation transfer functions identified by Moret-Bailly (CREIL) provide mechanisms for intrinsic redshifting in the extended photospheres of quasars and other active galactic nuclei which are consistent with the observed rate of cosmic attenuation. This thesis contends:

1) Radiation transfers occur naturally in low pressure gases with meta-stable states, intrinsically both redshifting and blueshifting spectra in a heterodyned construction of absorptions and stimulated emission.
2) The redshifting effect is greatly amplified in close proximity to high flux sources such as quasars, where the hot CREIL process varies the rate of redshifting as a function of flux, wavelength, pressure, ionization state and distance.
3) This bias causes the distance and magnitude of quasars and other brilliant objects to be greatly overestimated relative to galactic standard candles. Since the faces of quasars are used to detect Lyman absorptions, this also leads to underestimates of interstellar gas densities.
4) The hot CREIL zone about quasars also proportionally redshifts the quasar's host galaxy, and reddens objects observationally in line with quasars and other super bright objects.

5) Most of the reddening normally attributed to particle dust in the 'early' universe is due to a convergence in the near IR of radiation from many cosmic sources.

6) Obler's paradox, the failure of the microwave background to converge with the near - IR is the result of what is the equivalent of complete internal refraction at great distances.

7) The brilliance and polarity of objects in close proximity to each other are complimented by these spectral energy sharing effects.

8) The flux of our own galaxy contributes to reddening and redshifting, and increases the attenuation rate of objects we observe at very high redshifts. This changes the distance modulus, increasing attenuation with increasing distance.

9) These exponential functions are misrepresented by the relativistic distance modulus, which underestimates the net attenuation factor of space. This understated attenuation is misconstrued as an evolutionary effect, diminishing estimates of the true population distribution of galaxy types and morphologies.

10) Harmonic fluctuations in distant objects are evidence of these factors, artificially inducing an appearance of evolutionary peaks in the intensity and population of quasars.

11) The Tully-Fisher relationship exists because the spatial distribution of CREIL effects in galaxies convert the gross magnitude of the galaxy into broadened hydrogen alpha lines. This stimulates emissions in the galactic disk.

12) Gamma rays emitted from broad line galaxies are intrinsically redshifted.

13) CREIL causes the Wilson-Bappu effect and increases the coherency in light from the sun.

14) Primal matter is generated throughout space and time in galactic cores.

15) The cosmic microwave background is a byproduct of the stimulated emissions and radiation transfer functions responsible for our redshifted view of reality.

16) There is a correlating effect in the radiation transfer function of interstellar space permitting the kinetic flux from our own galaxy to impose a power attenuation factor on radiation from distant sources. This accelerated departure from celestial kinematics requires a mechanism which may also couple gravitational and electromagnetic forces at the quantum level through molecular Cooper effects in a Bose-Einstein condensation. This coupling nulls the zero-point energy field, providing a solution to Obler's paradox and MOND dynamics.

## Conclusion

The Wilson hypothesis should stand: The light curves of supernovae and quasars over many epochs must clearly demonstrate time dilation if the calculated rate of expansion is occurring. The failure of this hypothesis cries for an alternative cosmology either within or outside of current physical law boundaries. The convergence of the near IR continuum and the degeneration of the radio spectrum are powerful evidence of a celestial CREIL radiation transfer function. The speculations contained in this paper contend solutions are possible within the constraints of existing physical laws with a Maxwellian interpretation of gravity. Moret-Bailly has proposed a viable mechanism: Stimulated emission in meta-stable states in atomic and molecular hydrogen thermally coupled to the infrared and microwave background by elementary radiation transfer functions. The periodic symmetry of space about our own galaxy can best be explained by similar radiation transfers, stimulated by our own photons in the interstellar medium via Bose-Einstein condensations of discrete interstellar molecules.

Not every possible solution presented here is correct, but the mechanisms do offer viable alternatives to the constrained big bang model that are based on real observational data. The meticulous and copious data collected by thousands of dedicated scientists are telling us a story, and our best chance of properly interpreting the cosmos is found in the detailed spectra and light curves of supernovae, quasars and other brilliant objects.

**Predictions**
The SNAP program data should provide us a much better picture of SN Ia spectra and light curves at high red shift. Although we will still suffer from the lack of a complete local reference, this data may be definitive as to whether the current population of high redshift supernovae are binary representations of a SNe Ia and a *near* SNe Ia partners, or type SN Ic hypernovae. It will become statistically certain to what extent the light-curves of well-characterized supernova experience time dilation.

Laboratory scale determinations of the CREIL effect could be conducted between orbiting telescopes and the space shuttle. Using UV and IR lasers, measurable amounts of differential redshifting should be observable with a multi-wavelength Michleson Interferometers.

Since the 'hot' CREIL process displaces quasars with respect to most other objects in redshift distance estimates, the local concentration is significantly higher than current estimates that do not factor in exponential distance modulus effects. Proper motion of quasars will be confirmed.

Rejuvenating areas of light elements must occur throughout the universe. We will continue to observe widely varying levels of light and heavy element abundance into the distant past. The areas of light element abundance will provide the key to a qualified revision of entropy theory.

**Acknowledgements**
Extensive use has been made of the Los Alamos National Laboratory prepublication archives and other websites identified in the references. The comments and critic of Halton Arp, Ben Bromley, Terry Girard, Stacy McGaugh, Jon Middleditch, Jacque Moret-Bailly, Stacy Palen and Ed Wright have been greatly appreciated. My colleagues, Tim Doyle, Grant Porter, Mont Johnson, Jane Johnson, John Shipley, and Lee Pearson have been highly encouraging and supportive. I. L. Davis has been a patient proctor. The Weber State University Physics department has provided access to resource material. Tommy Highsmith provided invaluable advice on research strategy.

**Note 1**. The latest paper from the same researchers, (much to the chagrin of researchers who used the data to resurrect the cosmological constant); interprets the highest redshift supernovae as indicating the expansion rate then did an about face and the universe has started to contract (Riess 2004).

**Note 2.** Wang, the principle author of the 'Multicolor' method, has released a second paper with a convincing argument that peculiar supernova have a conic feature, possibly caused by a binary partner, which disrupts the explosion in the incipient stages. This conic intrusion redirects the mass flow and, depending upon the direction of the observer, the supernova peak and light curve length may be either enhanced or retarded. The Delta(15) value is determined during this period. The linear phase, which begins about twenty days after the light curve peaks, is used to calculate the Beta(BV). By then the disruptive flow has been dissipated and the nickel burning signature of an SN Ia is burning full bloom. This may be why there is little correlation between the two methods: The difference between the Delta(15) and Beta(BV) predictions for a given supernova is an indices of the degree of peculiarity. Since few of the median range redshift supernova use in this survey have been identified as 'peculiar' (an odd fact which also point towards a Malmquist effect), the lack of correlation is not invalidating.

To remove the time dilation factor from the Delta(15) values: (Delta(15) + 10) * z - 10 where z = $(1000*10^{cz})/C$, C being the speed of light and cz the encrypted z-shift used by Wang. This assumes the second derivative of the light curve slope is 33%. Varying the slope up to 50% produces the same trend.

**Note 3.** The Calan Tololo survey contains 18 supernova found at redshifts between z =0.014 and 0.101 (Parodi). A multiple linear regression analysis of the Calan Tololo survey indicates redshift is only responsible for as much as 9% of the variance in the Width ($MB_{(15)}$) of supernova light curves. Predicted width = 0.625 +0.18735* Magnitude -0.003893 x redshift. The Magnitude/light-curve width relationship correlates to a contribution of 28% of the variance. The accuracy is +/ -0.4 with 95% confidence at 2-sigma. (Allen). Since the magnitude/curve width contribution is three times greater, a distance bias of only 4%, a rough rule of thumb estimate for Malmquist effects, leaves no room for time dilation.

A significant percentage (~36%) of the local population is considered *peculiar*, demonstrating near normal Ia spectra but idiosyncratic light curves (Li). None of the currently observed population of high redshift supernovae are classified as *peculiar*. This is consistent with the hypothesis high redshift supernovae represent a most luminous homogenous supernova family: SNIc or hypernova.

Goldhaber, Sullivan and others have tested for, and excluded Malmquist biases from their studies. However, time dilation corrections have been made *a prior* to performing the bias test. In general, the care taken in these studies is extraordinary and most of the data impeccable. However this paper contends the time dilation correction is grossly overstated and complimentary, via the free parameter, to the Malmquist bias, therefore eliminating statistical detection of this bias. This is why free parameters are a poor choice over multi-variant iterative statistical analysis that in this case will yield contradictory results.

Some early supernovae expansion studies, which are based on photosphere deceleration, are independent of light-curve widths (Phillips 1999). However, they are still biased by a subtle magnitude correlation: As more brilliant supernova exploded, the matter is more widely distributed before the shockwave expansion is reeled in by gravity, the broader radial distribution slows the gravimetric deceleration rate.

### The K Corrections - SCP cross correlation method

To compare supernova spectra over many epochs, they are first normalized in both magnitude and spectral range to an arbitrary z-shift of 0.48:

$$K_{QR} = -2.5 Log_{10} \left[ \frac{1}{(1+z)} \right] \frac{\int dI_R I_R f_I(I_R) R(I_R) \int dI_e I_e g^Q I(I_e) Q(I_e)}{\int dI_R I_R g^R I(I_R) R(I_R) \int dI_e I_e f(1+z)(I_e) Q(I_e)}$$

General form of the cross-filter K corrections, term placement varies for emitter,
observer and reference frames selected (Hoggs).

After the spectra are transferred to a common reference epoch, each of the spectrum transformed from earlier or later epochs is compared to reference epoch templates. In the reference frame, the explosion curves are aligned by *matching the peak magnitude*. Then a stretch factor is selected which best aligns the explosion period of the time -shifted supernova with the light-curve of the same magnitude in the reference frame. Hubble expansion curves based on the calculated magnitude and the redshift are then used to estimate Hubble's constant, $H_o$. (Nugent et al., 2002, Goldhaber et al., 2001).

**Note 4.** This process is also apparent in the Sunyaev-Zel'dovich effect, which broadens the wavelengths of cosmic rays and correspondingly increases in the energy level of the microwave background. (Sandoval-Villabaso). Relativistic Compton scattering is governed by the Klein-Nishima Formula (Weisstein).

**Note 5.** This model is for relative comparison and ignores all other sources of attenuation. In this simulation, these broadband models were developed using Chandrasekhar's equations *before* Jacque's papers on CREIL were discovered. The CREIL mechanism perfectly matches the empirical model.

**Note 6.** Since CREIL intrinsically broadens the Hydrogen emission lines, the impinging of light from the galactic core should, during cosmic redshifting, cause a slight degeneracy in the calculated plane of the galaxy relative to our own. I was looking for published evidence of this effect, and found it in the excellent thesis on Tully-Fisher by Simard. In fact, most of the collaborating evidence gathered in this paper has been based on "If this is true then…" searches, which have been very rewarding to the author.

**Note 7.** Critics argue this self-sustaining universe is impossible because the Second Law of Thermal-dynamics rules out any solution that does not eventually convert all matter into useless thermal energy. But the Big Bang itself is a violation of this very law. Since the universe *does* exist, the Cosmological Principle requires a mechanism for energy reapportionment. Whether this mechanism is a big bang coupled to a tracter beam, quantum friability, or the slow decay mechanism proposed by Hawking is an open argument.

Jerry W. Jensen. Scientist, ATK Propulsion (independent research)
Jerry.Jensen@ATK.com, Fredbonyea@yahoo.com (comments welcome)

## References

Abraham, Bergh, Nair, "A New Approach to Galaxy Morphology: I, Analysis of the Sloan Digital Sky Survey Early Data Release", Astro-ph 0301239, Jan 2003.

Allen, Andrew, personal communication, Mar 2003.

Arny, Explorations in Stars, Galaxies and Planets, McGraw-Hill, 2002.

An & Sellgren "SPATIAL SEPARATION OF THE 3.29 µm EMISSION FEATURE AND ASSOCIATED 2 µm CONTINUUM IN NGC 7023", Astro-Ph 0307416, Jul 2003.

Anderson et al., "Study of the anomalous acceleration of Pioneer 10 and 11", Gr-Qc 0104064, Apr 2001

Arp, "Origins of Quasars and Galaxy Clusters", Astro-ph 0105325, May 2001.

Barton et al., "The Tully-Fisher Relation as a Measure of Luminosity Evolution: A Low Redshift Baseline for Evolving Galaxies", Astro-Ph 0011022, Jan 2000.

Barden et al., "Alpha Rotation Curves of z~1 Galaxies: Unraveling the Evolution of the Tully-Fisher Relation", Astro-ph 0302392, Feb. 2003.

Barris et al. "23 High Redshift Supernovae from the IfA Deep Survey: Doubling the SN Sample at z>0.7", Astro Ph 0310843

Basset & Kunz, "Is Cosmic Distance-Duality Violated", Astro Ph 0312443, Dec 2003.

Bell, M., "Discrete Intrinsic Redshifts from Quasars to Normal Galaxies", Astro-Ph 0211091, Nov 2002.

Bell & Comeau, "Further Evidence for Quantitized Intrinsic Redshifts in Galaxies: Is the Great Attractor a Myth?" Astro-ph 0305112, May 2003.

Benitez & Riess, "The magnification of SN 1997ff, the Farthest know Supernova", Astro-Ph 0207097.

Benitez et al., "Faint Galaxies in deep ACS observations", Astro-Ph 0309077, Sep 2003.

Bennetti et al., "Supernova 2002bo: Inadequacy of the Single Parameter Description", Astro-Ph 0309665

Best et al., "Red galaxy over-densities and the varied cluster environments of powerful radio sources with z~1.6", Astro-ph, 0304035, Apr 2003.

Bierman and Tanco, "Ultrahigh Energy Cosmic Rays Sources & Experimental Results", Astro-ph 0302299, Jan 2003.

Blaschke et al., "Relative Standard of Measurement and Supernova Data", Astro-ph 0302001, Feb 2003.

Bloom, Frail, Kulkarni, "GRB Energies and the GRB Hubble Diagram: Promises and Limitations", Astro-ph 0302210, Feb 2003.

Boller et al, "XMM-Newton observation of the ULIRG NGC 6240: The physical nature of the complex Fe K line emission", Astro-ph 0307326, July 2003, See also Hogg 2003.

Böhm et al., "The Tully-Fisher Relation at Intermediate Redshift", Astro-ph 030923, Sep 2003.

Bottema, "MOND Rotation Curves for Spiral Galaxies with Cepheid -Based Distances", Astro-ph 0207469, July 2002.

Boughn & Crittenden, "A Correlation of the Cosmic Microwave Sky with Large Scale Structure", Astro-ph 0305001, Apr 2003.

Burbidge et al., "QSOs Associated with Messier 82", Astro-ph 0303625, Mar 2003.

Bromley et al; "Density-Dependent Luminosity Functions for Galaxies in the Las Campanas Redshift Survey", Astro-Ph 9805197, May, 1998.

Breuck et al. A Statistical Study of Emission Lines from High Redshift Radio Galaxies, astro-ph/0008264

Cardini et al., "A study of the Mg II 2796.34 A emission line in late--type normal, and RS CVn stars", Astro-ph 0307296, Jul 2003.

Carlstrom , Joy et al., "Imaging the Sunyaev-Zel'dovich Effect", Astro-ph 9905255, May 1999.

Chandrasekhar, S, Radiation Transfer, Dover NY, NY, 1960.

Chartas, Brandt, & Gallagher, "XMM-NEWTON REVEALS THE QUASAR OUTFLOW IN PG1115+080", 0306125, Jun 2003.

Coil et al., "Optical Spectra of Type Ia Supernovae at z = 0.46 and z = 1.2", Astro Ph 0009102, Sep 2000.

Cline, Crotty & Lesgourgues, "Does the small CMB quadrupole moment suggest new physics? Astro Ph

0305020, May 2003.

Derouich et al., "On the collisional depolarization and transfer rates of spectral lines by atomic hydrogen. III: application to $f$-states of neutral atoms", Astro Ph 0307553, Jul 2003.

Dunlop, JS., "The Host Galaxies of Radio-Loud and Radio-Quiet Quasars", Astro-Ph 0103238, Mar 2001.

Evans et al., "The Clustering of Ultra-High Energy Cosmic Rays and Their Sources", Astro-ph 0212533, Dec 2002.

Floyd et al., "The host galaxies of luminous quasars", Astro-Ph 0308436, Aug 2003.

Floyd, "Luminous Quasars and Their Hosts; Accretion at the Limit?", Astro-ph 0303037, Mar 2003.

Foley et al., "Optical Photometry and Spectroscopy of the SN 1998bw-like Type Ic Supernova 2002ap", Astro-Ph 0307136, Jul 2003.

Frail et al., "Beaming in Gamma-ray Bursts: Evidence for a Standard Energy Reservoir", Astro Ph 0102282, Feb 2001.

Freedman et al., "Final Results from the Hubble Space Telescope Key Project to Measure the Hubble Constant 1", Astro-ph 0012376, Dec 2000.

Fynbo et al., "On the Ly Alpha Emission from Gamma-ray Burst Host Galaxies: Evidence for Low Metallicities", Astro-ph 0306403, Jun 2003.

Galama et al., "Discovery of the peculiar supernova 1998bw in the error box of GRB980425", Astro-ph 9806175, Jun 1998.

Garnavitch et al., "Constraints on Cosmological Models from Hubble Space Telescope Observations of High-z Supernovae", Astro-ph 970123, Jan 1997.

Genzel et al., "The Stellar Cusp Around the Supermassive Black Hole in the Galactic Center", Astro-ph 0305423, May, 2003.

Goldhaber et al., "Timescale Stretch Parameterization of Type Ia Supernova B-Band Light Curves", Astro-ph 0104382, Apr 2001.

Goldhaber et al., "Observation of Cosmological Time Dilation Using Type Ia Supernovae as Clocks", Astro-ph, Feb 1996.

Grebenev *et al.* 2000 *Science* 289 1532

Goto et al., "Evolution of the Colour--Radius Relation and the Morphology--Radius Relation in the SDSS Galaxy Clusters", Astro-Ph 0312040, Dec 2003

Hall et al., "Very Red and Extremely Red Galaxies in the Fields of z ~ 1.5 Radio-Loud Quasars", Astro-Ph 0101239, Jan 2001.

Hamuy et al., "The Absolute Luminosities of the Calan/Tololo Type Ia Supernovae", Astro-Ph 9609059, Sep 1996.

Hamuy et al., "The Distance to SN 1999cm from the Expanding Photosphere Method", Astro-ph 0105005, May 2001.

Hawkins, M.R.S., "Time Dilation and Quasar Variability", Astro-ph 0105073, May 2001.

Hikage et al., "Minkowski Functionals of SDSS galaxies I: Analysis of Excursion Sets", Astro-ph 0304455, Apr 2003.

Hoeflich et al., "Aspherical Explosion Models for SN1998bw/GRB 980425", Astro-Ph 9808086

Hodge & Castelaz, "Relationship Between Cepheid and Tully-Fisher Distance Calculations", Astro-ph 0304030 v4, Apr 2003.

Hodge & Castelaz 2003, "Neighboring Galaxies Influence on Rotational Curve Asymmetry", Astro-ph 0305022, May 2003.

Hogg & Blanton "The dependence on environment of the color-magnitude relation of galaxies", Astro-ph 0307336, Jul, 2003.

Hogg et al., "The K correction", Journal of Astro-Physics, Astro-Ph 0210394 v1, Oct 2002.

Inskip et al., "Deep spectroscopy of z~1 6C radio galaxies - II. Breaking the redshift-radio power degeneracy" Astro-ph 0208549

Iwamoto, "A `Hypernova' model for SN 1998bw associated with gamma-ray burst of 25 April 1998", Astro-Ph 9806382, Jun 1998.

Li & Filippenko, "Observations of Type Ia Supernovae, and Challenges for Cosmology", Astro-Ph 0310529, Oct 2003

Jain et al., "Large Scale Alignment of Optical Polarizations from Distant QSOs using Coordinate Invariant Statistics", Astro-ph 0301530, Jan 2003.

Jain et al., "Quasar-Galaxy And Galaxy-Galaxy Cross-Correlations: Model Predictions with Realistic Galaxies", Astro-ph 0304203 11, Apr 2003.

Jarvis et al., The mass of radio galaxies from low to high redshift, Astro ph 0112341, Matt

Jensen & Moret-Bailly, "Propagation of Electromagnetic Waves in Space Plasma" Astro Ph, 0401529

Karl et al., "Binaries Discovered by the SPY Project. III. HE2209-1444: a Massive, Short Period Double Degenerate", Astro-Ph 0309730.

Kasen, Nugent, Thomas, Wang, "Could There Be A Hole In Type Ia Supernova", Astro Ph 0311009.

Kishimoto, Antonucci, Blaes, "A First Close Look at the Balmer Edge Behavior of the Quasar Big Blue Bump", Astro-ph 0304118, Apr 2003.

Kukula et al., "A NICMOS imaging study of high-z quasar host galaxies", Astro-Ph 0010007, Jan 2000.

Lamb, Donaghy, Graziani, "Gamma-Ray Bursts as a Laboratory for the Study of Type Ic Supernovae", Astro Ph 0309463, Sep 2003.

Leibundgut & Suntzeff, "Optical Light Curves of Supernova", Astro-ph 0304112, Apr 2003.

Leibundgut et al., "Time Dilation in the Light Curve of the Distant Type Ia Supernova SN 1995K" Astro-Ph 9605134, 22 May 1996.

Li & Filippenko, "Observations of Type Ia Supernovae, and Challenges for Cosmology", Astro-Ph 0310529, Oct 2003

Lilly et al., "The Metallicity of Star-forming Gas of Cosmic Time", Astro-ph 0209243, Sep 2002.

Lilly, Carollo & Stockton, "The metallicities of star-forming galaxies at intermediate redshifts $0.47 < z < 0.92$", Astro-ph 0307300, Jul 2003.

Lopez-Corredoira & Gutierrez "The field surrounding NGC 7603: cosmological or non-cosmological redshifts?" Astro Ph 0401147, Jan 2004

Lubin & Sandage, "The Tolman Surface Brightness Test for the Reality of the Expansion. III. HST Profile and Surface Brightness Data for Early-Type Galaxies in Three High-Redshift Clusters" Astro-Ph 0106563, Jun 2001,

Lubin & Sandage, "The Tolman Surface Brightness Test for the Reality of the Expansion. IV. A Measurement of the Tolman Signal and the Luminosity Evolution of Early-Type Galaxies", Astro-Ph 016566, Jun 2001

Lyutikov, et al., "Polarization of prompt GRB emission: evidence for electromagnetically-dominated outflow", Astro-ph 0305410 May, 2003.

MacMillan, D,S, "Quasar Apparent Proper Motion Observed by Geodetic VLBI Networks", Astro-Ph 0309825, Sep 2003.

Matheson et al., "Photometry and Spectroscopy of GRB 030329 and Its Associated Supernova 2003dh: The First Two Months", Astro-Ph 0307435, Jul 2003.

Matteucci, "What Determines Galactic Evolution?", Astro- Ph 0210540, Oct 2002.

McDowell et al., "Chandra Observations of Extended X-ray Emission in Arp 220", Astro-ph 0303316.

Middleditch, John, "A White Dwarf Merger Paradigm for Supernova and Gamma -Ray Bursts", Astro Ph 0311484.

Milvang-Jensen et al., "The Tully-Fisher Relation of Cluster Spirals at z = 0.83", Astro-ph 0211598.

Miralda-Escude 1997 et al., "New Statistical Measures of the Lya Forest Spectra for Accurate Comparison to Theoretical Models", Astro-ph 9710230, 1997.

Miralda-Escude 2002 et al., "On the Evolution of the Ionizing Emissivity of Galaxies and Quasars Required by the Hydrogen Reionization", Astro-ph 0211071, Oct 2002.

Moiseev, "Two-dimensional spectroscopy of double-barred galaxies", Astro-ph 0211105, Nov 2002.

Moret-Bailly, Propagation of Light in Low Pressure Ionized and Atomic Hydrogen: Astrophysical Implications", Astro-ph 0305180 May, 2003.

Moret-Bailly[2], "Theory on the Quantification of Redshifts", Astro-ph 0307140, Jul 2003.

Moret-Bailly[3], "Computation of the Spectra of the Quasars", Astronomy & Astrophysics Sep 23, 2003

Narlikar, Introduction to Cosmology, Cambridge, 1993.

Nomato et al., "The Properties of Hypernovae: SNe Ic 1998bw, 1997ef, and SN IIn 1997cy", Astro-Ph 0003077, Mar 2002.

Nugent, Kim and Perlmutter, "K-corrections and Extinction Corrections for Type Ia Supernovae", Astro-Ph 0205351, May, 2002.

Nugent et al. , "Evidence for a Spectroscopic Sequence among Supernovae", Astro-ph 9510004, Oct 1995.

Oguri et al., "Observations and Theoretical Implications of the Large Separation Lensed Quasar SDSS J1004+4112", Astro Ph 0312429, Dec 2003.

Ostlie & Carroll, An Introduction to Modern Astrophysics, Addison Wesley Pub. Co., 1996.

Murphy, Curran & Webb, "A Search for High Redshift Molecular Absorption Lines toward

Millimetre-Loud Optically Faint Quasars", Astro-ph 0303074, 4 Mar 2003.

Pankaj et al., "Large Scale Alignment of Optically Polarizations from Distant QSOs using Coordinate Invariant Statistics", Astro-ph 0301530, 27 Jan 2003.

Peebles, P. J. E., <u>Principles of Physical Cosmology</u>, Princeton, 1993; also Astro-ph 0209403, Astro-ph 0208037, Astro-ph 0207347, Astro-ph 0309269.

Perlmutter & Schmidt, "Measuring Cosmology with Supernovae", Astro-ph 0303428, Mar 2003.

Perlmutter et al., "Supernovae, Dark Energy, and the Acceleration Universe", Physics Today, Apr 2003.

Perlmutter et al., "Discovery of a Supernova Explosion at Half the Age of the Universe", Astro-Ph 9712212, Dec 1997.

Perlmutter 1998 et al., "Measurements of Omega and Lambda from 42 High-Redshift Supernovae", Astro-Ph/9812133, Dec 1998.

Phillips et al., "Infrared Light Curves of Type Ia Supernovae", Astro-ph 0211100, Nov 2002.

Phillips 1999 et al., "The Reddening-Free Decline Rate versus Luminosity Relationship for Type Ia Supernovae", Astro-ph 9907052, Jul 1999.

Phillips 1998 et al., "Observational Evidence from Supernovae for an Acceleration Universe and a Cosmological Constant", Astro-Ph 9805201, May 1998.

Prigara, "Einstein's coefficients and the nature of thermal blackbody radiation", Astro-ph 0211355.

Prochaska et al., "The ESI/Keck II Damped Lya Abundance Database", Astro-ph. 0304312, May 2003.

Prochaska, "The Chemical Uniformity of High Z DLA Protogalaxies", Astro Ph 0209193, Sep 2002

Ralston et al., "The Virgo Alignment Puzzle in the Propagation of Radiation on Cosmology", Astro Ph 0311430, Nov 2003

Raunch, "The Lyman Alpha Forest in the Spectra of QSOS", Annual Review Astronomy, Astro-phys 36:267-316, 1998, http://nedwww.ipac.caltech.edu/level5/Sept01/Rauch/frames.html.

Ray & Das, "Tolman-Bayin type Static Charged Fluid Spheres in General Relativity", Astro Ph 0401005.

Richardson et al., "A Comparative Study of the Absolute-Magnitude Distributions of Supernovae", Astro-Ph/0112051, Dec 2001.

Reichards et al., "A Catalog of Broad Absorption Line Quasars from the Sloan Digital Sky Survey Early Data Release", Astro-ph 0301019, Jan 2003.

Richards et al., "Red and Reddened Quasars in the Sloan Digital Sky Survey", Astro-ph 0305305.

Riess et al., "Type Ia Supernova Discoveries at z>1 From the Hubble Space Telescope: Evidence for Past Deceleration and Constraints on Dark Energy Evolution", Astro-Ph 0402512, Feb 2004

Riess et al., "Observational Evidence from Supernovae for an Acceleration Universe and a Cosmological Constant", Astro-Ph 9805201, May 1998.

Riess et al., Time Dilation from Spectral Feature Age Measurements of Type Ia Supernovae", Astro-ph 9707260, Jul 1997.

Sako, Masao, "CNO Line Radiation and the X-ray Bowen Fluorescence Mechanism in Optically-Thick, -Highly-Ionized Media", Astro-ph 0305514, May, 2003.

Sandoval-Villabaso, and Garcia-Colin, "Notes on the Sunyaev-Zel'dovich thermal effect", Astro-ph 0305144 May 2003.

Sandoval-Villabaso and Garcia-Colin, "A convolution Integral Representation of the Thermal Sunyaev-Zel'dovich Effect", Astro-ph 0208440, Feb 2003.

Sandoval-Villabaso, and Garcia-Colin, "The Thermal and Kinamatic Sunyaev-Zel'dovich effects Revisited", Astro-Ph 0310465, Oct. 2003.

Sanders & Mirable, "LUMINOUS INFRARED GALAXIES", Astro Ph 749-792, 1996

Saxanov, et al., "Quasars: the Characteristic Spectrum and the Induced Radiative Heating", Astro ph 0305233, May 2003.

SCP (Supernova Cosmology Project): http://techreports.larc.nasa.gov/ltrs/ltrs.html.Shimasaku, "Statistical Properties of Bright Galaxies in the SDSS Photometric System", Astro-ph 0105401, May 2001

Shimasaku, "Statistical Properties of Bright Galaxies in the SDSS Photometric System", Astro-Ph 0105401, May 2001.

Schombert, "the Evolution of Galaxies,: A Metaphysics Viewpoint", Astro-Ph 0403005, Mar 2004.

Simard et al., "The Magnitude-Size Relation of Galaxies out to z ~ 1", Astro-ph 9902147, Feb 1999.

Simard, Luc, "The Local Tully-Fisher Relation and Morphological Dependence of the Tully-Fisher Relation", http://www.ucolick.org/~simard/phd/root/node48.html, *Sep 1996*

Stritzinger et al., "Optical Photometry of the Type Ia SN 1999ee and the Type Ib/c SN 1999ex in IC 5179", Astro-ph 0206438, Jun 2002.

Strolger et al. "The Type Ia Supernova 1999aw: A Probable 1999aa Like Event in a Low-Luminosity Host Galaxy", Astro-Ph 0207409v1 18 Jul 2002.

Stanev, Todr, "On the Luminosity of the Ultra High Energy Cosmic Ray Sources", Astro-ph 0303123.

Sullivan et al., "The Hubble Diagram of Type Ia Supernovae as a Function of Host Galaxy Morphology", Aston Ph 0211444, Nov 2002.

Teerikorpi et al., "Observational Selection Bias affecting the Determination of the Extragalactic Distance Scale", Annual Review of Astronomy and Astrophysics, Vol. 35, 1997.

Tonry et al., "Cosmological Results from High Z Supernovae SNe Ia", Astro-ph 0305008, May 03.

Treyer & Wanbsganss, "Astrometric Microlensing of Quasars", Astro Ph 0311519

Van Putten et el., "Observational Evidence for a correlation between Peak-luminosities and Beaming in GRBs", Astro-ph 0307058, Jul 2003.

Vink et al., "SLOW TEMPERATURE EQUILIBRATION BEHIND THE SHOCK FRONT OF SN 1006", Astro-ph 0303051, Mar 2003.

Wang, "Supernova Explosions: Lessons from Spectropolarimetry", Astro Ph 0311299

Wang, Goldhaber, Aldering, Perlmutter, "Multi-Color Light Curves of Type Ia Supernovae on the Color-Magnitude Diagram: a Novel Step Toward More Precise Distance and Extinction Estimates", Astro-ph 0302341, Feb 2003.

Weidner & Sells, "Elementary Modern Physics", Allyn and Bacon, Boston 1980.

Weinstein & Akhoury, "Quantitized Cosmology II: De Sitter Space", Astro Ph 0312249, Dec 2003

Williams et al., "Imaging and Demography of the Host Galaxies of High-Redshift Type Ia Supernovae", Astro Ph 0310432, Oct 2003.

WMAP First Year Wilkinson Microwave Anisotropy Probe Observations, Series of Articles in Astro-ph, released February 13[th], 2003.

Wold et al., " The Surface Density of Extremely Red Objects in High-z Quasar Fields", Astro-ph 0303090, Mar 2003.

Wolf et al., "The Combo-17 Survey: Evolution of the Galaxy Luminosity Function from 25,000 Galaxies with $0.2 < z < 1.2$", Astro-ph 0208345, Oct 2002.

Wolf et al., "The Evolution of Faint AGN Between $z = 1$ and $z= 5$ from the COMBO-17 Survey", Astro-ph 0304072, Apr 2003.

Wolf et al., "Tests of Lorentz Invariance using a Microwave Resonator: An Update", Astro-ph 0306047

Wright, "Theoretical Overview of Cosmic Microwave Background Anisotropy", Astro-ph 0305591Yan, "Constraining evolution in the Halo Model using galaxy redshift surveys", Astro-ph 0307248

Yoshiaki, "Galactic Center Shells and a Recurrent Starburst Model", Astro Ph 0303328, Mar 2003.

Zackrisson et al., "Can Microlensing Explain the Long-Term Optical Variability of Quasars?" Astro-ph 0306434, Jun 2003.

Ziegler et al., "Internal kinematics of spiral galaxies in distant clusters Part I", Astro-ph 0309267

Zubko, Dwek & Arendt, "New Interstellar Dust Models Consistent with Extinction, Emission, and Abundance Constraints", Astro Ph 0312641, NASA Goddard zubko@stars.gsfc.nasa.gov

**Websites**

Castles, Quasar Lensing, http://cfa-www.harvard.edu/glensdata/

National Space Science and Technology Center, Huntsville AL  ken-ichi.nishikawa@nsstc.nasa.gov

NASA, Chandra X-ray Observatory, http://chandra.nasa.gov/

NASA/IPAC http://nedwww.ipac.caltech.edu

Langley Technical Report Server:http://techreports.larc.nasa.gov/ltrs/ltrs.html

Los Alamos National Archives: http://lanl.arxiv.org/

Michelson-Morley hyper physics.phy-astro.gsu.edu/hbase/relativ/morley/html

McGaugh, The MOND Pages, http://www.astro.umd.edu/~ssm/mond/

Peebles, http://pupgg.princeton.edu/www/jh/research/peebles_james.htmlx

Raunch, "The Lyman Alpha Forest in the Spectra of QSOS", Astro-ph 36:267-316, 1998, http://nedwww.ipac.caltech.edu/level5/Sept01/Rauch/frames.html

Search engine:  http://www.dogpile.com/index.gsp

Sloan Digital Sky Survey, http://www.sdss.org/

Supernova Cosmology Project (SCP): http://techreports.larc.nasa.gov/ltrs/ltrs.html

Weisstein, Eric W., http://scienceworld.wolfram.com/physics/Klein-NishimaFormula.html

Wright, Edward., "Cosmological Fads and Fallacies", www.astro.ucla.edu/~wright/cosmology.